\documentclass[smallcondensed]{svjour3}     
\usepackage{array}
\usepackage{amsmath}
\usepackage{color}
\usepackage{amssymb}
\usepackage{booktabs}
\usepackage{graphicx}
\usepackage{caption}
\usepackage{subcaption}
\usepackage{url}
\usepackage{natbib}
\usepackage{pifont}
\newcolumntype{H}{>{\setbox0=\hbox\bgroup}c<{\egroup}@{}}

\usepackage{multirow}
\usepackage{paralist}
\usepackage{verbatim}

\usepackage{xfrac}

 \makeatletter
\def\bbordermatrix#1{\begingroup \m@th
  \@tempdima 2\p@
  \setbox\z@\vbox{%
    \def\cr{\crcr\noalign{\kern\p@\global\let\cr\endline}}%
    \ialign{$##$\hfil\kern\p@\kern\@tempdima&\thinspace\hfil$##$\hfil
      &&\;\hfil$##$\hfil\vspace{-0.5pt}\crcr
      \omit\strut\hfil\crcr\noalign{\kern-\baselineskip}%
      #1\crcr\omit\strut\cr}}%
  \setbox\tw@\vbox{\unvcopy\z@\global\setbox\@ne\lastbox}%
  \setbox\tw@\hbox{\unhbox\@ne\unskip\global\setbox\@ne\lastbox}%
  \setbox\tw@\hbox{$\kern\wd\@ne\kern-\@tempdima\left[\kern-\wd\@ne
    \global\setbox\@ne\vbox{\box\@ne\kern\p@}%
    \vcenter{\kern-\ht\@ne\unvbox\z@\kern-\baselineskip}\,\right]$}%
  \null\;\vbox{\kern\ht\@ne\box\tw@}\endgroup}
\newcommand\mmsubsection{\@startsection{subsection}{3}{\z@}%
                                     {-3.25ex\@plus -1ex \@minus -.2ex}%
                                     {-1.5ex \@plus -.2ex}
                                     {\normalfont\normalsize\bfseries}}
  
 \makeatother

\newcommand{\TODO}[1]{}
\newcommand{\reffig}[1]{Figure~\ref{#1}}
\newcommand{\refthe}[1]{Theorem~\ref{#1}}
\newcommand{\reffoot}[1]{Footnote~\ref{#1}}
\newcommand{\reftab}[1]{Table~\ref{#1}}
\newcommand{\refeqb}[1]{Equation~\ref{#1}}
\newcommand{\refdef}[1]{Definition~\ref{#1}}
\newcommand{\refsec}[1]{Section~\ref{#1}}
\newcommand{\refap}[1]{Appendix~\ref{#1}}
\newcommand{\refidd}[1]{Identity~\ref{#1}}
\newcommand{\refco}[1]{Corollary~\ref{#1}}

\newcommand{\proofOf}[1]{\hfill\mmsubsection{Proof of #1:}}

\begin{document}
\renewcommand{\cite}{\citep}

\title{Generalization of Clustering Agreements and Distances
}
\subtitle{for Overlapping Clusters and Network Communities}
\author{Reihaneh Rabbany 
 \and Osmar R. Za\"{\i}ane }%

\date{Received: date / Accepted: date}

\authorrunning{Rabbany et al.}
\institute{	Department of Computing Science, 
University of Alberta\\
             \email{\{rabbanyk
             ,zaiane\}@ualberta.ca}        
}

\maketitle

\begin{abstract}

A measure of distance between two clusterings has important applications, including  clustering validation and ensemble clustering. Generally, such distance measure provides navigation through the space of possible clusterings. 
Mostly used in cluster validation, a normalized clustering distance, a.k.a. agreement measure, compares a given clustering result against the ground-truth clustering. Clustering agreement measures are often classified into two families of pair-counting and information theoretic measures, with the widely-used representatives of Adjusted Rand Index (ARI) and Normalized Mutual Information (NMI), respectively.
This paper sheds light on the relation between these two families through a generalization. 
It further presents an alternative algebraic formulation for these agreement measures which incorporates an intuitive clustering distance, which is defined based on the analogous between cluster overlaps and co-memberships of nodes in clusters. 
Unlike the original measures, it is easily extendable for different cases, including overlapping clusters and clusters of inter-related data for complex networks. 
These two extensions are, in particular, important in the context of finding clusters in social and information networks, a.k.a communities.


\end{abstract}
\section{Introduction} 
A cluster distance, accordance, similarity, or divergence has different applications. \emph{Cluster validation} is the most common usage of cluster distance measures. In particular, in external evaluation, a clustering algorithm is validated on a set of benchmark datasets by comparing the similarity of its results against the ground-truth clusterings
. 
Another notable application is \emph{ensemble, or consensus Clustering}, where results of different clustering algorithms on the same dataset are aggregated. A notion of distance between alternative clusterings is used in modeling and formulating this aggregation, i.e. to find a clustering that has the minimum average distance to the alternative clusterings\footnote{
Refer to \citet{aggarwal2014cluseringbook}, Chapter 23 on clustering validation measures (in particular the section on external clustering validation measures); and Chapter 22 on cluster ensembles (in particular the section on measuring similarity between clustering solutions).}.
Another closely related application is multi-view clustering \cite{Cui2007multiv}, where the objective is to find different clusterings of the same dataset, which are usually in different sub-spaces of the data, and could represent different views of that dataset. In the same context, one might be interested to find the sub-spaces that result in different/similar clusterings. 

Clustering distance measures are well-studied and widely-used in cluster validation, where a normalized distance measure is used to average the performance of an algorithm over different datasets, and to compare different algorithms. Some of the most widely used clustering agreement measures are: Adjusted Rand Index (ARI), Normalized Mutual Information (NMI), and Variation of Information (VI). 

In this paper, we first study the well-known clustering agreement measures, which are classified into two families of pair counting and information theoretic measures. 
Then we highlight the relation between these two families by presenting a generalized formula that covers both. 
Next, we elaborate on the limitations of these measures in handling inter-related data-points, and also overlapping clusters. 
These two limitations are in particular problematic when measuring distance between clusterings in the context of information networks. 

Networks encode the relationship between data-points, and clusters on a real network are known to be overlapping. 
Many methods for network clustering, a.k.a. community mining, have been proposed in recent years; the reader could refer to \citet{Fortunato10survey} for a survey. In the evaluation and comparison of these algorithms, often the classical clustering agreement measures, mostly NMI, are applied. Here, we discuss the effect of neglecting relations between data points, e.g. edges in networks, in measuring communities distance, and derive extensions of our generalized formula to incorporate such relationships.  

We further discuss the difficulty of extending the current contingency (a.k.a. overlap, or confusion) based formulation for the general cases of overlapping clusters. We tackle this by presenting an alternative algebraic formulation for a clustering distance, based on the analogous relationship of cluster overlaps and co-memberships of nodes. From the proposed algebraic formulation we could derive the original formulations, and we could also easily derive new forms that are appropriate for the cases of overlapping clusters, and also network clusters.
\section{Clustering Agreement Measures: Short Survey}
\label{sec:agreements}
Consider a dataset $D$ consisting of $n$ data items, $D = \{d_1,d_2,d_3\dots d_n\}$. A partitioning $U$ partitions $D$ into $k$ mutually disjoint subsets, $U=\{U_1,U_2\ldots U_{k}\}$; where 
\(
D = \cup_{i=1}^{k} U_i \; \text{and} \; U_i\cap{U_j}=\emptyset
\; \forall i\neq j
\).
%
There are several measures defined to examine the similarity, a.k.a agreement, between two partitioning of the same dataset. More formally, let $V$ denote another partitioning of the dataset $D$, $V=\{V_1,V_2\ldots V_{r}\}$. 
Clustering agreement measures are originally introduced based on counting the pairs of data items that are in the same/different partition in $U$ and $V$. 
Each pair $(d_i,d_j)$ of data items is classified into one of four groups based on their co-memberships in $U$ and $V$; which results in the following pair-counts.

 \newcommand{\hlinee}{\hline}
\begin{table}[h]
\centering 
\begin{tabular}{l | c| c}
~ & Same in $V$ & Different in $V$ \\
\hlinee
Same in $U$ & $\quad M_{11} = TP \quad $ & $\quad M_{10} = FP  \quad$ \\
\hlinee
Different in $U$ & $\quad M_{01} = FN  \quad$ & $\quad M_{00} = TN  \quad$ \\
\end{tabular}\vspace{-10pt}
\end{table}

Here, $M_{11}$/$M_{00}$ counts the number of pairs that are in the same/different partitions in both $U$ and $V$. 
$M_{10}$/$M_{01}$ sums up those that belong to the same/different partitions in $U$ but are in different same/partitions according to $V$. 
Note that $M_{11} + M_{00} + M_{10} + M_{01} = \binom{n}{2}$. 
When one of these partitionings, for instance $V$, is the true partitioning i.e. the ground-truth, these could also be referred to as the 
true/false positive/negative scores, denoted by TP, FP, TN, and FN in the table\footnote{Also denoted by $a$, $b$, $c$, $d$ letters for the notational convenience in some literature.}. 

These pair counts 
are often derived using the following contingency table a.k.a. confusion table \cite{Hubert85ARI}. 
The contingency table is a $k \times r$ matrix of all the possible overlaps between each pair of clusters in $U$ and $V$, where its $ij$th element shows the intersection of cluster $U_i$ and $V_j$, i.e. $n_{ij} = |U_i \cap V_j|$. 


\begin{table}[h]
\centering\vspace{-10pt}
\begin{tabular}{c|c c c c |c}
~ & $\quad V_1$ & $V_2$ & $\dots$ & $V_r \quad$ & marginal sums\\
\hlinee 
$U_1$ & $\quad n_{11}$ & $n_{12}$ & $\dots$ & $n_{1r}\quad $ & $n_{1.}$\\
$U_2$ & $\quad n_{21}$ & $n_{22}$ & $\dots$ & $n_{2r}\quad $ & $n_{2.}$\\
 $\vdots$ &  $\quad \vdots$ &  $\vdots$ &  $\ddots$ &  $\vdots\quad $ &  $\vdots$\\
$U_k$ & $\quad n_{k1}$ & $n_{k2}$ & $\dots$ & $n_{kr}\quad $ & $n_{k.}$\\
\hlinee
marginal sums & $\quad n_{.1}$ & $n_{.2}$ & $\dots$ & $n_{.r}\quad $ & $ n $ \\
\end{tabular}
\end{table}

The last row and column show the marginal sums of $n_{i.}=\sum_j n_{ij}$, and $n_{.j}=\sum_i n_{ij}$, where in this case of disjoint clusters we also have $n_{i.} = |U_i|$, and $n_{.j}=|V_j|$.
The pair counts can then be computed using the following formulae.
 \begin{align*}
M_{10} &= \sum_{i=1}^k\binom{n_{i.}}{2} - \sum_{i=1}^k\sum_{j=1}^r\binom{n_{ij}}{2}  ,\quad
M_{01} =  \sum_{j=1}^r \binom{n_{.j}}{2} - \sum_{i=1}^k\sum_{j=1}^r\binom{n_{ij}}{2} \\
M_{11} &=  \sum_{i=1}^k\sum_{j=1}^r \binom{n_{ij}}{2},\quad
M_{00} =  \binom{n}{2} + \sum_{i=1}^k\sum_{j=1}^r\binom{n_{ij}}{2}  - \sum_{i=1}^k\binom{n_{i.}}{2} - \sum_{j=1}^r\binom{n_{.j}}{2}
 \end{align*}
These \emph{pair counts} have been used to define a variety of different clustering agreement measures \cite{Manning08IRbook}. Here, we briefly explain the most common measures; the reader can refer to \citet{Albatineh06} for a complete survey.

Considering co-membership of data points in the same or different clusters as a binary variable, 
\emph{Jaccard} agreement between 
clustering $U$ and $V$ can be defined as 
 \(
  J = \sfrac{TP}{(FP + FN + TP)} = \sfrac{M_{11}}{(M_{01} + M_{10} + M_{11})} 
 \).
\emph{Rand Index} is defined similarly to Jaccard, but it also values pairs that belong to different clusters in both partitionings, i.e. true negatives: 
\( RI = \sfrac{(M_{11} + M_{00}) }{(M_{11} + M_{01} + M_{10} + M_{00})}\), which gives: 
\vspace{-8pt}
\begin{equation} RI = 1+ \frac{1}{n^2-n}(2\sum_{i=1}^k\sum_{j=1}^r n_{ij}^2 - (\sum_{i=1}^k n_{i.}^2 + \sum_{j=1}^r n_{.j}^2))
\label{eq:RI}
\end{equation}
The \emph{Mirkin Index} is a transformation of Rand Index, defined as $n(n-1)(RI-1)$, which is equivalent to $RI$ when comparing partitionings of the same dataset~\cite{Wu09KDD}.
 %
 \emph{F-measure} is a weighted mean of the precision ($P$) and recall ($R$),
\(F_{\beta} = \frac{({\beta}^2 + 1) PR}{ {\beta}^2 P + R} \) where \(P = \sfrac{M_{11}}{(M_{11}+M_{10})}\) and  \(R = \sfrac{M_{11}}{(M_{11}+M_{01})}\).
The parameter $\beta$ indicates how much recall is more important than precision. 
The two common values for $\beta$ are $2$ and $.5$; the former weighs recall higher than precision while the latter favours the precision more. 
%


There is also a family of  \emph{information theoretic} based measures. 
These measures consider the overlaps between clusters in $U$ and $V$, as a joint distribution of two random variables, i.e. the cluster memberships in $U$ and $V$. 
The entropy of cluster $U$, $H(U)$, the joint entropy of $U$ and $V$, $H(U,V)$, their mutual information, $I(U,V)$, and their\emph{Variation of Information}~\cite{Meilq07VI}, $VI(U,V)$ are then defined as: 
\vspace{-8pt}
 \begin{align*}
&H(U)=-\sum_{i=1}^k \frac{n_{i.}}{n}\log(\frac{n_{i.}}{n}), 
\quad H(V) = - \sum_{j=1}^r \frac{n_{.j}}{n}\log(\frac{n_{.j}}{n})\\[-3mm]
&H(U,V)=-\sum_{i=1}^k\sum_{j=1}^r\frac{n_{ij}}{n}\log(\frac{n_{ij}}{n}), \quad I(U,V) = \sum_{i=1}^k\sum_{j=1}^r\frac{n_{ij}}{n}\log(\frac{n_{ij}/n}{n_{i.}n_{.j}/n^2}) 
\end{align*}\vspace{-8pt}
\begin{equation}
\label{eq:vi}
VI(U,V)= \sum_{i=1}^k\sum_{j=1}^r\frac{n_{ij}}{n}\log(\frac{n_{i.}n_{.j}/n^2}{n_{ij}^2/n^2}) 
\end{equation}

All the pair counting measures defined here, except Mirkin, have a fixed range of $[0,1]$. 
The above information theoretic measures, however, do not have a fixed range. For example, the mutual information ranges between $(0, log k]$, and the range for variation of information is $[0, 2 \log \max (k,r)]$ \cite{Wu09KDD}.
Having a fixed range, i.e. being {\em{normalized}}, is a desired property for partitioning agreement indexes, since we often require to compare/average agreements 
over different datasets. 
%
%
Consequently, normalized variations of mutual information are defined 
\cite{Vinh10AMI}.
%
%
The most commonly used normalization forms are:
\begin{equation}
\label{eq:nmi}
NMI_{\Sigma} =  \frac{2I(U,V)}{H(U)+H(V)} 
\quad \text{and}\quad NMI_{\sqrt{}} = \frac{I(U,V)}{\sqrt{H(U)H(V)}}\vspace{-3pt}
\end{equation}
%
%

Beside having a fixed range, a clustering agreement measure should also have a \emph{constant baseline}  \cite{Vinh10AMI,Hubert85ARI}.
As an example, consider the case where agreement between a clustering and the ground-truth is measured as $0.7$. 
If the baseline of the measure is not constant, it can be $0.6$ in one settings and $0.2$ in another, then this $0.7$ value can be both a strong or a weak agreement.
 %
 \emph{Correction for chance} is adjusting a measure to have a constant (usually $0$) expected value for agreements no better than random. 
 This adjustment is calculated based on an upper bound on the measure, $Max[M]$, and its expected value, $E[M]$, as: 
 \begin{equation}
  AM = \frac{M - E[M] }{ Max[M] - E[M]} \label{eq:adjust}
 \end{equation}
The \emph{Adjusted Rand Index (ARI)} is proposed in \cite{Hubert85ARI}, assuming that the contingency table is constructed randomly when the marginals are fixed, i.e. the size of the clusters in $U$ and $V$ are fixed. With this assumption, $RI$ is a linear transformation of $\sum_{i,j}\binom{n_{ij}}{2}$, and 
$  E\left(\sum_{i,j}\binom{n_{ij}}{2}\right) = \sum_{i}\binom{n_{i.}}{2} \sum_{j}\binom{n_{.j}}{2} / \binom{n}{2}$.
Hence, adjusting $RI$ with upper bound $1$ results in the following formula: 
\begin{align}
\label{eq:ari}
  ARI = \frac{\sum\limits_{i=1}^k\sum\limits_{j=1}^r\binom{n_{ij}}{2} - \sum\limits_{i=1}^k\binom{n_{i.}}{2} \sum\limits_{j=1}^r\binom{n_{.j}}{2} / \binom{n}{2} }
  {\frac{1}{2}[\sum\limits_{i=1}^k\binom{n_{i.}}{2} + \sum\limits_{j=1}^r\binom{n_{.j}}{2} ] - \sum\limits_{i=1}^k\binom{n_{i.}}{2} \sum\limits_{j=1}^r\binom{n_{.j}}{2} / \binom{n}{2} } 
 \end{align}
There is also an approximate formulation \cite{Hubert85ARI,Albatineh06} for this expectation defined as $  E(\sum_{i,j} n_{ij}^2) = \sum_{i} n_{i.}^2 \sum_{j} n_{.j}^2 / n^2$, which results in a slightly different formula for the $ARI$, i.e.
  \begin{equation}
  \label{eq:ariapprox}
  ARI' = \frac{\sum\limits_{i=1}^k\sum\limits_{j=1}^r{n_{ij}}^{2} - \sum\limits_{i=1}^k {n_{i.}}^{2} \sum\limits_{j=1}^r {n_{.j}}^{2} /  {n}^{2} }
  {\frac{1}{2}[\sum\limits_{i=1}^k {n_{i.}}^{2} + \sum\limits_{j=1}^r {n_{.j}}^{2} ] - \sum\limits_{i=1}^k {n_{i.}}^{2} \sum\limits_{j=1}^r {n_{.j}}^{2} /  {n}^{2} } 
  \end{equation}

There are several variations of pair counting agreement measures, 
such as Gamma, Hubert, Pearson, etc. 
These measures, however, become similar or even equivalent after correction for chance.  
More specifically, \citet{Albatineh06} show that many of these measures are linear transformations of $\sum_{i,j} n_{ij}^2$, 
 i.e. each measure could be written as 
 \( \alpha + \beta \sum_{i,j} n_{ij}^2\), 
 where $\alpha$ and $\beta$ depend on the marginal counts, $n_{i.}$ or $n_{.j}$, but not on the $n_{ij}$. For example for the Rand Index we have:
 \(\alpha = 1 - \frac{1}{n(n-1)} (\sum_{i} n_{i.}^2  + \sum_{j} n_{.j}^2)\), and \(\beta = 2/n(n-1)\).
They further prove that these measures become equivalent if their $\frac{1-\alpha}{\beta}$ ratio is the same, since their corrected for chance formula will all be as:
 \[
  \frac{{ \sum_{i,j} n_{ij}^2 - E(\sum_{i,j}  n_{ij}^2 )   }}
  { { \sfrac{1-\alpha}{\beta} -  E(\sum_{i,j}  n_{ij}^2 )    }}
 \]
\citet{warrens08} extended these results and included the inter-rater reliability indices from \emph{statistics}. 
Using the $2\times 2$ pair counting table, he has shown that all the pair counting clustering agreement measures after correction for chance become equivalent to one of the statistical inter-rater agreement indices. 
The well-studied \emph{inter-rater agreement indices} in statistics are defined to measure the agreement between different coders, rankers, or judges on categorizing the same data. 
Examples are the goodness of fit: chi-square test, the likelihood chi-square, kappa measure of agreement, Fisher's exact test, Krippendroff's alpha, etc. (see test 16 in \cite{statbook}). 
These statistical tests are also defined based on the contingency table which displays the multivariate frequency distribution of the (categorical) variables. 
%
Specifically, \emph{Cohen's kappa} is one the most widely used inter-rater agreement index; a chance corrected index of association defined for accessing the agreement between two raters, who categorize data into $k$ categories (defined as  
\( \kappa = \sfrac{\sum_{i=j}^k n_{ij} - \sum_{i=j}^k {E_{ij}}}{n - \sum_{i=j}^k {E_{ij}}}\) where \(E_{ij} = \frac{n_{i.}n_{.j}}{n} \)).
The equivalence of Cohen's kappa and the $ARI$ is proved by \citet{warrens08kappa}.  

\citet{Vinh09AMI} proposed the correction for chance of the information theoretic measures, and showed that Adjusted Variation of Information ($AVI$) is equivalent to \emph{Adjusted Mutual Information} ($AMI$).     
They derived the expected value of the mutual information assuming the sizes of the clusters are fixed, i.e. similar to the $ARI$'s assumption on the hypergeometric model of randomness. 
In more details, the expected value is defined as:
\newcommand{\oma}{\mathfrak{m}}
\[\scalebox{0.99}{$
 E[I(U,V)]=  \sum_{i,j} \sum\limits_{\oma=\max(n_{i.}+n_{.j}-n,1)}^{min(n_{i.},n_{.j})}
\frac{\oma}{n}\log(\frac{n\oma}{n_{i.}n_{.j}}) 
\frac{n_{i.}!n_{.j}!(n-n_{i.})!(n-n_{.j})!}{n!\oma!(n_{i.}-\oma)!(n_{.j}-\oma)!(n-n_{i.}-n_{.j}+\oma)!}$
}\]
From which, $I$ can be adjusted for chance using~\refeqb{eq:adjust}: 
\(AMI = \frac{I - E[I]}{Max[I]- E[I])}\); 
 where $ Max[I]$ is one of upper bounds on $I$:
\[
\scalebox{0.91}
{$I(U,V)\leq\min(H(U),H(V))\leq\sqrt{H(U)H(V)} \leq\frac{H(U)+H(V)}{2} \leq\max(H(U),H(V)) \leq H(U,V)$}
\]
The $AVI=AMI$ is true when the $1/2(H(U)+H(V))$ upper bound is used in the adjustment.
The formulation of $AMI$ includes big factorials, therefore is computationally complex, and less practical when compared to the $ARI$. 
\section{Generalization of Clustering Agreement Measures}
\label{sec:generalization}
In this section, we highlight the connection between pair counting and information theoretic measures, through defining a generalized formula that covers both. 
We start by noting the relation between the Rand Index ($RI$), as a representative of the pair counting measures, and the Variation of Information ($VI$),  as a representative for the information theoretic measures. 

\begin{proposition}
\label{th:varen}
VI (RI) of two partitionings is proportional to the conditional entropies (variances) of memberships in them (see \refap{app:proVIAI} for proof), i.e.  
\[VI(U,V)  =  {H(U|V)+H(V|U)} \quad \text{and} \quad
RI(U,V)  \propto  Var(U|V) + Var(V|U) 
\]
\end{proposition}
%
This proposition inspires defining a generalized distance for clusterings as:
\newcommand{\GAM}{\mathcal{D}}
\newcommand{\GAMP}[2]{\GAM_{#1}^{#2}}
\begin{definition}{Generalized Clustering Distance ($\GAM$)}
\label{def:gam}
 \vspace{-5pt}
 \begin{equation*}
\GAM_{\varphi}^{\eta}(U,V) =  \GAM_{\varphi}^\eta (U||V) + \GAM_{\varphi}^\eta (V||U), \quad
 \GAM_{\varphi}^\eta (U||V) = \sum_{v\in V} \left[   \varphi(\sum_{u\in U} \eta_{uv}) -\sum_{u\in U} \varphi(\eta_{uv})\right]
 \vspace{-3pt}
 \end{equation*}
 where $\eta_{uv}$ quantifies the similarity between the two clusters of $u\in U$ and $v \in V$, i.e. $\eta : 2^V\times 2^U \rightarrow \mathbb{R}$; and $\varphi:\mathbb{R} \rightarrow \mathbb{R}$.
\end{definition}
\begin{corollary}
\label{co:norm}
$\GAM$ is bounded if $\varphi$ is a positive superadditive function 
(proof in \refap{app:norm}), i.e. 
 \vspace{-5pt}
\[\varphi(x)\geq 0  \; \land \;
   \varphi(x+y) \geq  \: \varphi (x) + \varphi(y)  
  \implies  0 \leq \GAM_{\varphi}^\eta(U||V) \leq\, \varphi(\sum_{v\in V} \sum_{u\in U} \eta_{uv})
\]
\end{corollary}
\vspace{-3pt}
Using this bound as a normalizing factor, we define: 
\newtheorem{identity}{Identity}
\newcommand{\NGAM}{\mathcal{N}\GAM}
\begin{definition}{Normalized Generalized Clustering Distance ($\NGAM$)}
\label{def:ngam}
 \begin{align*}
 \NGAM_{\varphi}^\eta (U,V) = \frac{\GAM_{\varphi}^\eta (U,V)}{NF(U,V)} , \quad 
 NF(U,V) = {\varphi(\sum_{v\in V} \sum_{u\in U} \eta_{uv}) }
 \end{align*}
\end{definition}\vspace{-3pt}
We can show that the following two identities hold for the proposed $\NGAM$.
\begin{identity}
\label{id:vi} The Variation of Information (\refeqb{eq:vi}) derives from $\NGAM$ if we set $\varphi(x) = x\log x$, and $\eta$ as the overlap size: $\eta_{uv}=|u \cap v|$ (proof in \refap{app:idVI}), i.e.
\[\NGAM_{x\log x}^{|\cap|}(U,V) \equiv \frac{VI(U,V)}{\log n} \]
\end{identity}
\begin{identity}
\label{id:ri} The Rand Index (\refeqb{eq:RI}) derives from $\NGAM$ if we set $\varphi(x) = \binom{x}{2}$, and $\eta$ as the overlap size (proof in \refap{app:idRI}), i.e. 
\[\NGAM_{\binom{x}{2}}^{|\cap|}(U,V) \equiv 1-RI(U,V) \]  
\end{identity}

Similar to the \refidd{id:ri}, in the rest of this paper we consider clustering agreement~($\mathcal{I}$) and normalized distance~($\NGAM$) interchangeably using $\mathcal{I}=1-\NGAM$.
 \newpage
\newcommand{\AG}{\mathcal{A}\GAM}  
\newcommand{\AGAM}{\mathcal{A}\GAM}  

We further adjust the generalized distance to return the maximum of one, if $U$ and $V$ are independent.  
Assume the joint probability distribution $P_{U,V}(u,v) = \sfrac{\eta_{uv}}{\sum_{uv} \eta_{uv}}$, 
with the marginal probabilities of 
$P_{U}(u) = \sum_v P_{U,V}(u,v) = \sfrac{\eta_{.v}}{\sum_{uv} \eta_{uv}}$ and $P_{V}(v) = \sfrac{\eta_{u.}}{\sum_{uv} \eta_{uv}}$.
Then the independence condition for $U$ and $V$, $P_{U,V}(u,v) = P_U(u)P_V(v)$, translates into $\eta_{uv}= \sfrac{\eta_{u.}\eta_{.v}}{\sum_{uv} \eta_{uv}}$.
On the other hand, we have 
$ \GAM_{\varphi}^\eta (U,V) = \sum_{v\in V} \varphi(\eta_{.v}) + \sum_{u\in U} \varphi(\eta_{u.}) -2 \sum_{v\in V} \sum_{u\in U} \varphi(\eta_{uv})$, hence we define:
\begin{definition}{Adjusted Generalized Clustering Distance ($\AGAM$)}
\label{def:agam}
\vspace{-7pt}
\begin{align*}{
\AGAM_{\varphi}^\eta = \frac{\GAM_{\varphi}^\eta (U,V)}
{NF(U,V)}, \quad NF={ \sum_{v\in V} \varphi(\eta_{.v}) + \sum_{u\in U} \varphi(\eta_{u.}) - 2 \sum_{u\in U}\sum_{v\in V} \varphi \left(\frac{\eta_{.v}\eta_{u.}}{\sum\limits_{u\in U}\sum\limits_{v\in V} \eta_{uv} }\right) }}
\end{align*} 
\end{definition}
\begin{identity}
\label{id:nmi}
The Normalized Mutual Information (\refeqb{eq:nmi}) derives from $\AGAM$, if we set $\varphi(x) = xlogx$, and $\eta$ as the overlap size: $\eta_{uv}=|u \cap v|$ (proof in \refap{app:idNMI}), i.e.
\[\AG_{xlogx}^{|\cap|}(U,V) \equiv 1- NMI_{sum}(U,V) \] 
\end{identity}
\begin{identity}
\label{id:ari}
The Adjusted Rand Index of \refeqb{eq:ari} and \refeqb{eq:ariapprox} derive from $\AGAM$, if we set $\varphi(x)=x(x-1)$ and $\varphi(x)=x^2$ respectively, where $\eta$ is the overlap size, (proof in \refap{app:idNMI}), i.e.
\[\AG_{x^2}^{|\cap|}(U,V) \equiv 1- ARI'(U,V), \quad \AG_{x(x-1)}^{|\cap|}(U,V) \equiv 1- ARI(U,V) \] 
\end{identity}
This line of generalization is similar to the works in Bergman Divergence and $f$-divergences. 
For example, the mutual information and variance are proved to be special cases of Bergman information \cite{Banerjee2005}.
The (reverse) KL divergence and Pearson $\chi^2$ are shown to be $f$-divergences when the generator is $x\log x$ and $(x-1)^2$ respectively \cite{Nielsen2013}. 
Beside this analogy, our generalized measure is different from these divergences. One could consider our proposed measure as an (adjusted normalized) conditional Bergman entropy for clusterings. 
This relation is however non-trivial and is out of scope of this paper.  
 \vspace{-5pt}
\subsection{Extension for Inter-related Data}
\label{sec:intergam}
\vspace{-3pt}
All the agreement measures presented so far only consider memberships of data-points, and ignore any relations between them. Ignoring these relations is however problematic, as also mentioned by a few previous works. For example \citet{zhou2005new} discuss the issue of ignoring the distances between data-points when comparing clusterings, and propose to compare clusterings using a measure that incorporates the distances between representatives of the clusters.

\begin{figure}[t] 
\centering 
\includegraphics[width=.9\textwidth]{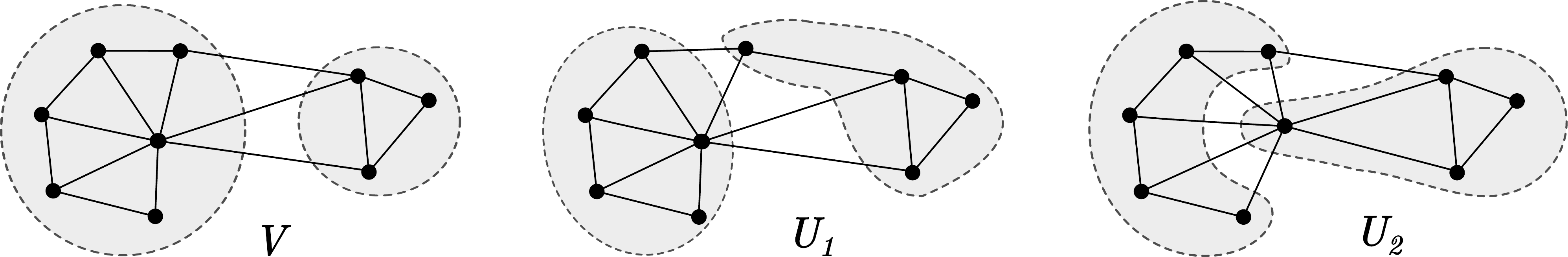}
\caption{\footnotesize 
Partitioning $U_1$ and $U2$ of the same graph with true partitioning $V$. 
Considering only the number of nodes in the overlaps, $U_1$ and $U_2$ have the same contingency table with $V$, i.e. $|\cap|(U_1,V)=|\cap|(U_2,V)=\{\{5,0\},\{1,3\}\}$. 
Therefore they have the same agreement with $V$, regardless of the choice of the agreement measure: $ARI$, $NMI$, etc. 
However if considering the edges, $U_1$ is more similar to the true partitioning $V$. This could be enforced using an alternative overlap function that incorporates edges, such as the degree weighted overlap function
: $\Sigma d(U_1,V) = \{\{18,0\},\{3,9\}\}$ and $\Sigma d(U_2,V) = \{\{14,0\},\{7,9\}\}$; or the edge based variation
: $\xi(U_1,V) = \{\{7,0\},\{0,3\}\}$ and $\xi(U_2,V) = \{\{4,0\},\{0,3\}\}$. }
\label{fig:exm}
\end{figure}

The extension of the clustering agreement or distance measures to incorporate the structure of the data, is in particular important when comparing \emph{clusterings of nodes within information networks}. 
An information network encodes relationships between data points, and a clustering on such network forms sub-graphs. 
Using the original clustering agreement measures to compare there clusterings, we only consider the nodes in measuring the clustering distance. 
It is however relevant that one should also consider edges when comparing two sub-graphs. \reffig{fig:exm} presents a clarifying example for the effect of considering or neglecting edges in comparing the network clusterings, a.k.a. communities.  

To incorporate the structure in our generalized distance measure, we can modify the overlap function $\eta$ in Definition 1. The overlap function from which the original $RI$ or $VI$ derive can be written as $|\cap|:\eta=\sum_{i\in u \cap v} 1$. Therefore the first intuitive modification to incorporate the structure is to consider a degree weighted function as:
 \vspace {-5pt}
\begin{equation*}
\label{eq:dwo}
\Sigma d : \eta_{uv}=  \sum_{i\in u \cap v} d_i
\end{equation*}
Using this $\eta$, well-connected nodes with higher degree weigh more in the distance. 
Alternatively, any other ranking criteria can be used depending on the underlying application. 
%
%
Another possibility is to alter $\eta$ to directly assess the structural similarity of these sub-graphs by counting their common edges: 
%
\vspace {-5pt}
\begin{equation*}
\label{eq:ebo}
\xi: \eta_{uv}= \sum_{i,j \in u \cap v} A_{ij}
\end{equation*}
%
One can consider many other alternatives for measuring overlap of two sub-graphs based on the application at hand. 
%
%
We revisit and delve deeper in this topic in \refsec{sec:AlgextNet}, after providing an alternative formulation for the clustering distance or agreement measures. 

 \vspace{-5pt}
\subsection{Extension for Overlapping Clusters}
\begin{figure}
\centering 
\begin{subfigure}[b]{0.45\textwidth}
\centering
\includegraphics[width=.85\textwidth]{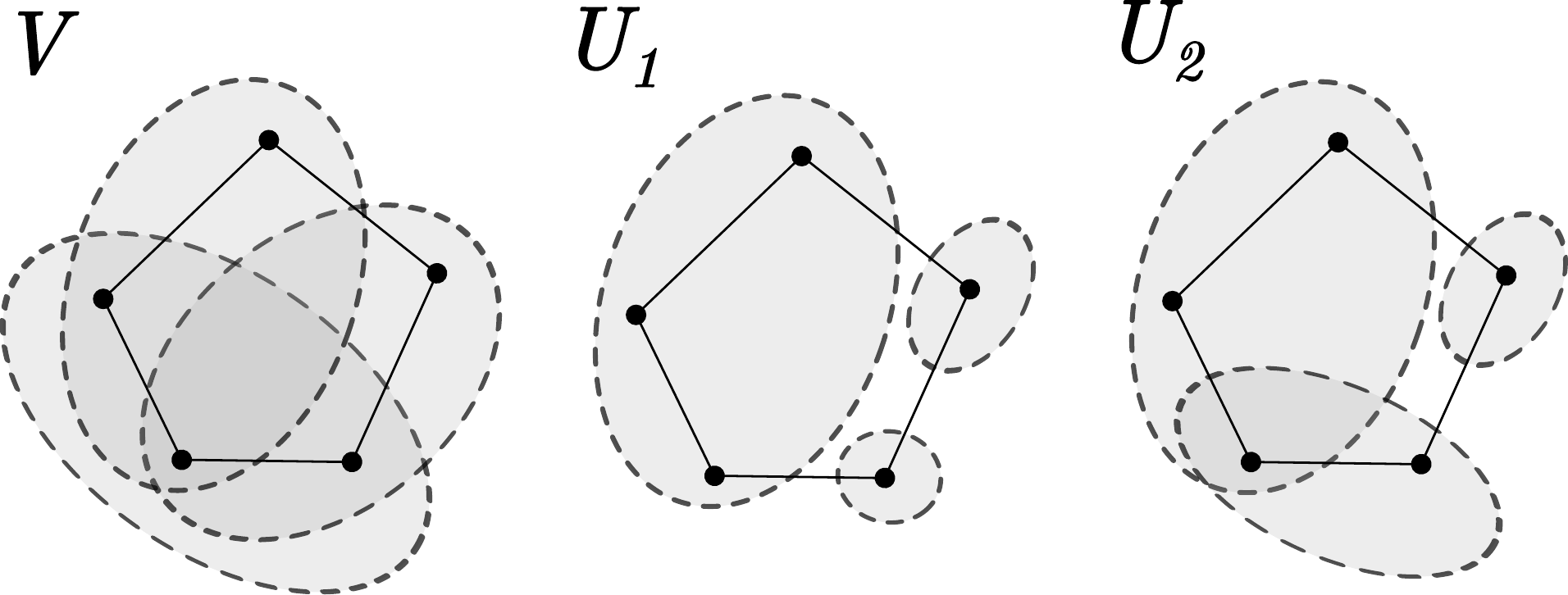} 
\caption{Omega index example: the pair-counts matrix of $U_1$ and $U_2$ with V are respectively $\{\{3,0,0\},\{1,1,1\},\{2,0,1\}\}$, and $\{\{3,0,0\},\{3,2,0\},\{0,2,0\}\}$. In more details, the second row of the latter matrix states that for the pairs of nodes that are clustered together in one cluster in $V$, we have $3$ of them in no clusters together in $U_2$, whereas $2$ are clustered once together.}  
\label{fig:omegaexm}
\end{subfigure}
\hspace{10pt}
\begin{subfigure}[b]{0.45\textwidth}\centering
\includegraphics[width=1\textwidth]{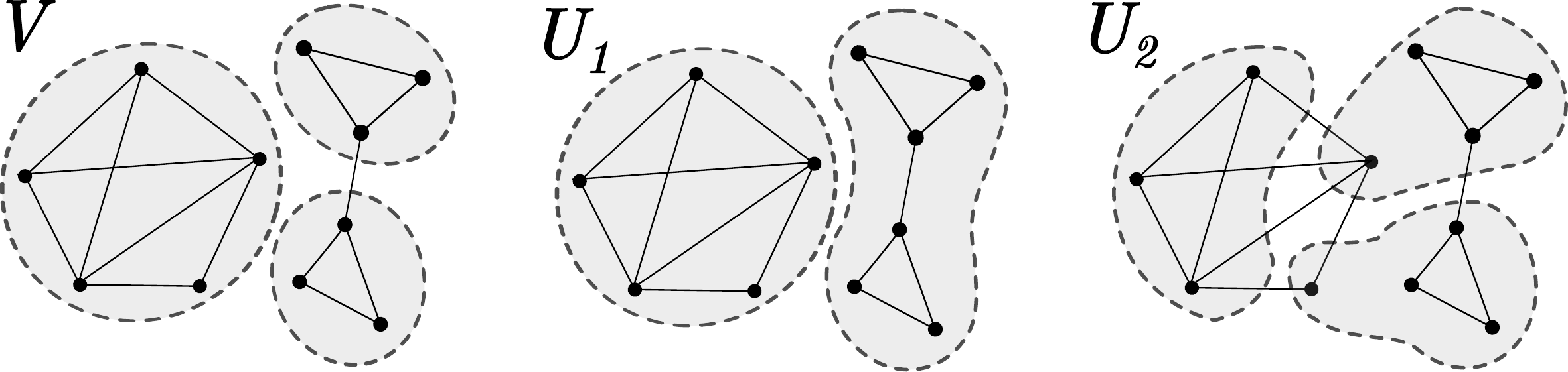}
\caption{Matching example: using the original formulation we have $NMI(U_1,V)$ = $0.78$ and $NMI(U_2,V)$ = $0.71$, whereas the overlapping version results in $NMI'(U_1,V)$ = $0.61$ and $NMI'(U_2,V)$ = $0.62$ with \citet{Lancichinetti08overlap} version and $NMI''(U_1,V)$ = $0.53$ and $NMI''(U_2,V)$ = $0.61$ with \citet{mcdaid2011normalized} version. }
\label{fig:setmatchingexm}
\end{subfigure}
\caption{\footnotesize Example for \textbf{Omega index} on the left: the pair-counts table for $U_1$ and $U_2$ with V have the same trace, and therefore they have the same degree of agreement with $V$ according to the Omega index. Example for \textbf{the problem of matching} on the right: using the set matching based measures, such as the overlapping version of the $NMI$, clustering $U_2$ is in higher agreement with $V$, while the non-overlapping version of $NMI$ suggests the opposite. Here we used a disjoint example to be able to compare the results quantitatively with the original $NMI$, this problem is however intrinsic to all the matching based measures, regardless of the overlapping or disjoint. }
\label{fig:overlappingExamples}
 \vspace{-10pt}
\end{figure}

  \vspace{-3pt}
\label{sec:overlappSur}
There are several non-trivial extensions of the clustering agreement measures for the crisp overlapping clusters \cite{collins1988omega,Lancichinetti08overlap,Xie2013}. 
Notably, \citet{collins1988omega} proposed the \textbf{Omega index} as a generalization of the (adjusted) rand index for non-disjoint clusters with crisp memberships. 
The Omega index expands the $2\times 2$ pair-counts table of $U$ and $V$, $\{\{M_{00},M_{10}\},\{M_{01},M_{11}\}\}$; so that $M_{ij}$ 
counts the pair of data points that appeared together in $i$ clusters of $U$ and $j$ clusters of $V$. 
Similar to the $RI$, trace of this matrix, i.e. $\sum_{i}
 M_{ii}$, is considered as the agreement index, which is further adjusted for chance using marginals of $M$. 
The Omega index reduces to the $(A)RI$ if the clusterings are disjoint. 
It however has a fundamental problem as it 
 only considers the pairs that appeared in the \emph{exact} same number of clusters together. For example, consider a pair of data points which are in $2$ clusters together in the ground-truth. The Omega agreement of a clustering that puts that pair together in $1$ cluster is the same as another clustering that puts them together in no clusters. Figure~\ref{fig:omegaexm} provides an illustrated example for such a case. 

Another commonly used measure for overlapping clusters \cite{gregory2010finding,Xie2013} is the extension of $NMI$ proposed by \citet{Lancichinetti08overlap}. 
The proposed measure does not reduce to the original $NMI$ if the clusterings are disjoint. 
This extension assumes a matching between clusters in $U$ and $V$, and only considers the best pair of clusters (with minimum conditional entropy) in the agreement calculation. 
A similar idea is also used in computing agreement between disjoint clusters, which is the basis of the set matching measures. 
These measures are known to suffer from the ``problem of matching'' \cite{Meilq07VI}. See Figure~\ref{fig:setmatchingexm} for a visualized example. 
The same problem exists with any of the agreement indexes that consider only the best matching, e.g. Balanced Error Rate with alignment, average $F1$ score, and Recall measures used in \cite{yang2013overlapping,mcauley2012discovering,mcdaid2011normalized}.
%
There is also a line of work on extensions for fuzzy clusters with soft membership \cite{Brouwer2008,Quere2010,campello2010generalized,Anderson2010,Hullermeier2012}. 
The fuzzy mesures, however, are not applicable to cases where a data point could fully belong to more than one cluster, i.e. crisp overlapping (such as example of \reffig{fig:exOv}) which are common in network clustering. The bonding concept presented by \citet{Brouwer2008} is similar to the main idea behind our extension for overlapping cases, which we discuss further in \refsec{sec:overlappAlgebra}.

The extension of the proposed $\GAM$  formula (Definitions \ref{def:gam}, \ref{def:ngam}, and \ref{def:agam}) for overlapping clusters is not straightforward. 
The $(\mathcal{A}/\mathcal{N})\GAM$ formula is indeed bounded for overlapping clusters, and reduces to the original formulation if we have disjoint covering clusters. 
However, the current formulation is not appropriate for comparing overlapping clusters, since it treats overlaps as variations and penalizes them. 
Consider an extreme example when we are comparing two identical clusterings, and therefore we should have $(\mathcal{A}/\mathcal{N})\GAM =0$ (i.e. the perfect agreement); this is true if there is no overlapping nodes, however as the number of overlapping nodes increase, $(\mathcal{A}/\mathcal{N})\GAM$ also increase (i.e. the agreement decreases). 

The difficulty of computing the agreement of different clusterings, and in particular their extension for general cases such as overlapping clusters, partly comes from the fact that there is no matching between the clusters from the two clusterings. 
Therefore, one should consider all the permutations, or only consider the best matching, which is cursed with the ``problem of matching'' as discussed earlier.
We overcome this difficulty by an alternative algebraic formulation for the clustering agreement measures, which takes the permutation out of the equation. 

\section{Algebraic Formulation for Clustering Distance}\label{sec:overlappAlgebra} 
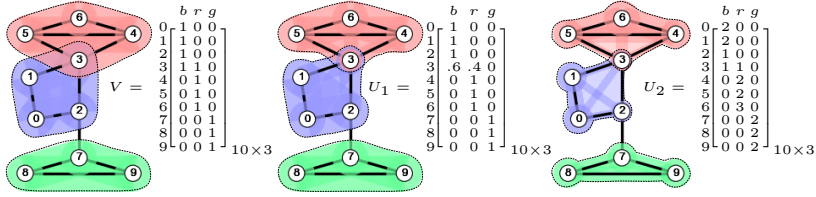
\begin{figure}
\vspace{10pt}
\hspace{10pt}
\begin{picture}(200,70)
\put(10,0){\includegraphics[width=.75\textwidth]{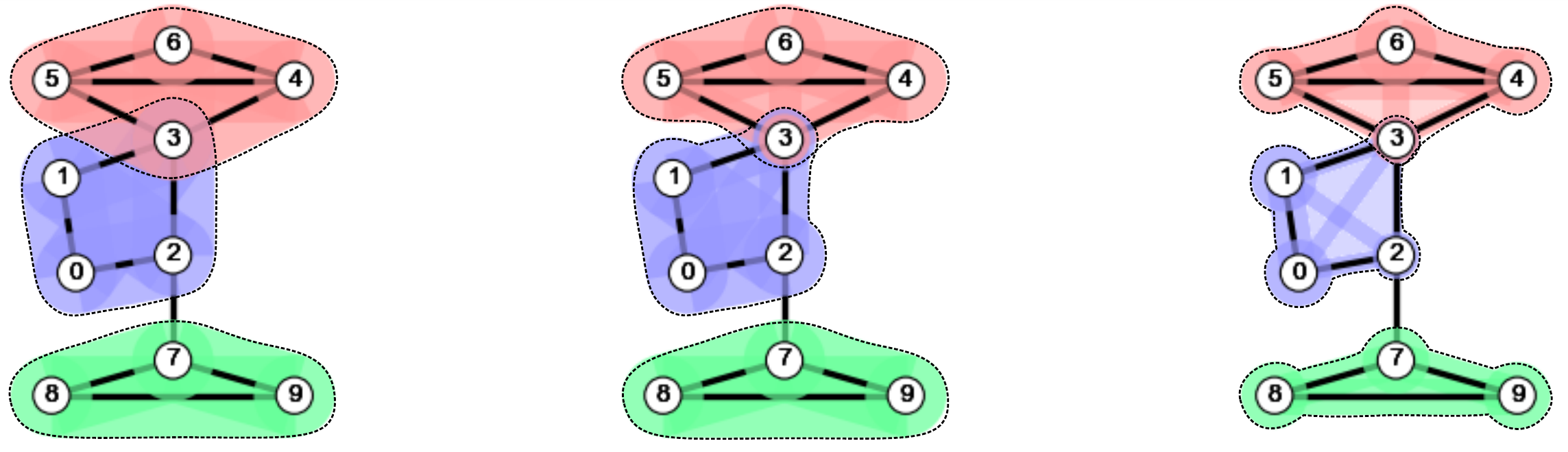}}
\put(50,40){\tiny$V=\bbordermatrix{~&b&r&g\cr
0& 1 & 0 & 0\cr
1&1 & 0 & 0\cr
2&1 & 0 & 0\cr
3&1 & 1 & 0\cr
4&0 & 1 & 0\cr
5&0 & 1 & 0\cr
6&0 & 1 & 0\cr
7&0 & 0 & 1\cr
8&0 & 0 & 1\cr
9&0 & 0 & 1}_{10\times3}$}
\put(147,40){\tiny$U_1=\bbordermatrix{~&b&r&g\cr 0& 1 & 0 & 0\cr 1 &1 & 0 & 0\cr 2 &1 & 0 & 0\cr 3 &.6 & .4 & 0\cr 4 &0 & 1 & 0\cr 5 &0 & 1 & 0\cr 6 &0 & 1 & 0\cr 7 &0 & 0 & 1\cr 8 &0 & 0 & 1\cr 9 &0 & 0 & 1}_{10\times3}$}
\put(250,40){\tiny$U_2=\bbordermatrix{~&b&r&g\cr0 & 2 & 0 & 0\cr  1&2 & 0 & 0\cr 2 &1 & 0 & 0\cr 3  &1 & 1 & 0\cr 4 &0 & 2 & 0\cr 5 &0 & 2 & 0\cr 6 &0 & 3 & 0\cr 7 &0 & 0 & 2\cr 8 &0 & 0 & 2\cr 9 &0 & 0 & 2}_{10\times3}$}
\end{picture}
\caption{\footnotesize Example of general matrix representation for a clustering: $V$ and $U_1$ are the classic overlapping clusters with crisp, and soft memberships respectively. Node $3$ fully belongs to both blue and red clusters in $V$, wherein $U_1$, it belongs $60\%$
 to the blue cluster and $40\%$ 
 to the red cluster. 
This representation is general in a sense that it could encode membership of nodes to clusters in any form, with no assumptions on the matrix, such as in $U_2$.}
\label{fig:exOv}
 \end{figure}
Let $U_{n \times k}$ denote a general representation for a clustering 
of a dataset with $n$ datapoints, i.e. $u_{ik}$ represents the memberships of node $i$ in the $k^{th}$ cluster of $U$.
Different constraints on this representation derive different cases of clustering.
For crisp clusters (a.k.a strict membership), $u_{ik}$ is restricted to ${0,1}$ ($1$ if node $i$ belongs to cluster $k$ and $0$ otherwise); whereas for probabilistic clusters (or soft membership), $u_{ik}$ could be any real number in $[0,1]$; see \reffig{fig:exOv} for examples. 
Fuzzy clusters usually assume an additional constraint that the total membership of a datapoint is equal to one, i.e. $u_{i.} = \sum_k u_{ik} = 1$. 
Which should also be true for disjoint clusters, as each datapoint can only belong to one cluster. 
 %

Here we first show that the clustering agreement measures discussed before can be reformulated in terms of this matrix representation. 
The size of overlaps between clusters in $U_{d\times k}$ and $V_{d\times r}$ --their contingency matrix-- derives as:
\[N= (U^TV)_{k \times r} = (V^TU)^T_{k \times r}\]

The agreement between disjoint clustering $U$ and $V$ is then calculated based on this contingency table.
More specifically, we can reformulate $\GAM$ and $\NGAM$ as:  
$$\GAM_{\varphi} = \left[
\mathbf{1} \varphi(N \mathbf{1}^T )  -\mathbf{1} \varphi(N) \mathbf{1}^T   \right]+ \left[
  \varphi(\mathbf{1} N) \mathbf{1}^T -\mathbf{1} \varphi(N) \mathbf{1}^T \right] ,\quad \NGAM_\varphi = \frac{\GAM_{\varphi}}{\varphi(\mathbf{1} N \mathbf{1}^T) }  $$
where $\mathbf{1}$ is a vector of ones with appropriate shape so that the matrix-vector product is valid, i.e. $\mathbf{1}N = [n_{.1}, n_{.2}, \dots n_{.r}]$ 
, and $N\mathbf{1}^T  = [n_{1.}, n_{2.}, \dots n_{k.}]^T$;  and $\varphi$ is applied element-wise to the given matix.
We can show that similar to the \refidd{id:vi} and \ref{id:ri}, the normalized Variation of Information derives from $\varphi (x) = x\log x $; and with $\varphi(x)= \binom{x}{2}$, $1-\NGAM_{\varphi}$ is equivalent to the rand index.  

Similarly, $\AG$ can be  reformulated as:\vspace{-5pt}
\[\AG_{\varphi} =\frac{\GAM_{\varphi}}{\frac{1 }{2 }  [\mathbf{1} \varphi(N \mathbf{1}^T ) +  \varphi(\mathbf{1} N) \mathbf{1}^T ]-E} ,\quad \; E =  \mathbf{1}  \varphi(  \frac{(N  \mathbf{1}^T )\times (\mathbf{1} N )}{\mathbf{1}  N \mathbf{1}^T}) \mathbf{1}^T \]

\begin{figure}[t]
\begin{picture}(60,100)
\put(15,35){\includegraphics[width=.38\textwidth]{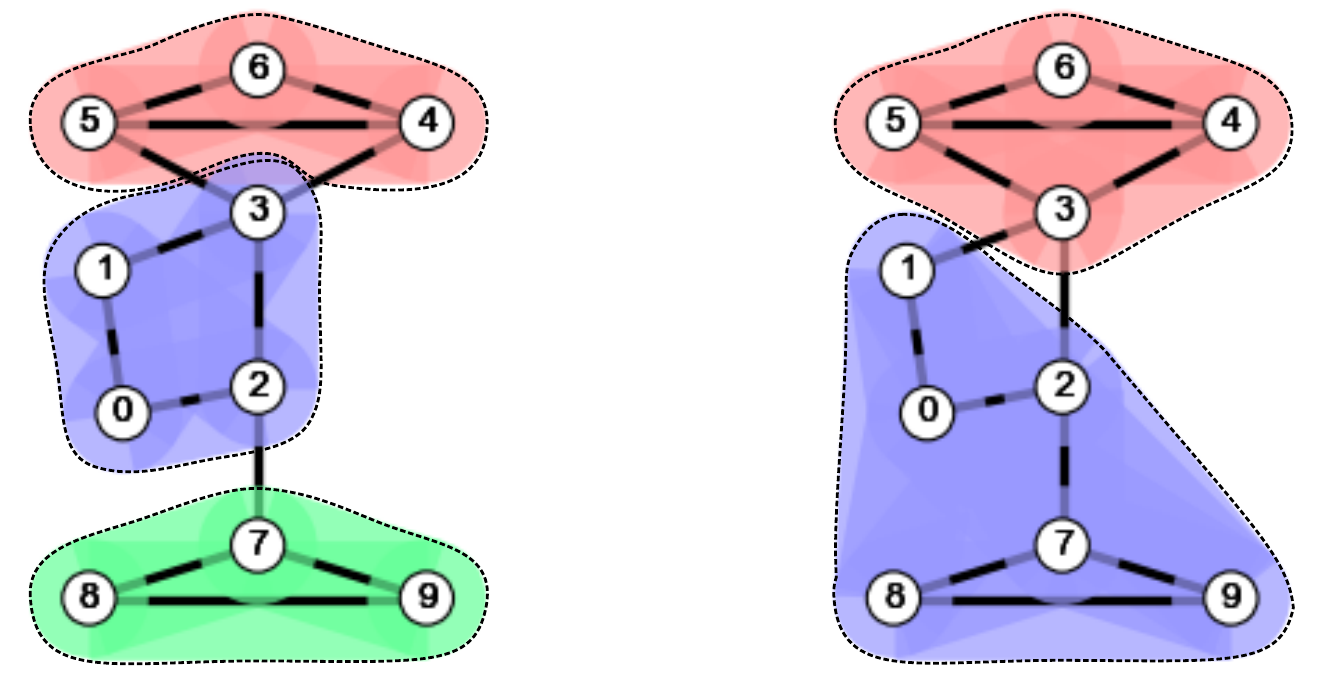}}
\put(48,65){\tiny 
$V =\bbordermatrix{~&b&r&g\cr
0& 1 & 0 & 0\cr  
1&1 & 0 & 0\cr  
2&1 & 0 & 0\cr  
3&1 & 0 & 0\cr  
4&0 & 1 & 0\cr  
5&0 & 1 & 0\cr  
6&0 & 1 & 0\cr  
7&0 & 0 & 1\cr  
8&0 & 0 & 1\cr  
9&0 & 0 & 1}$}
\put(132,65){\tiny $U = \bbordermatrix{~&b&r\cr
 0& 1 & 0\cr 1 &1 & 0\cr 2 &1 & 0\cr  3 &0 & 1\cr 4 &0 & 1\cr 5 &0 & 1\cr 6 &0 & 1\cr 7 &1 & 0\cr 8 &1 & 0\cr 9 &1 & 0}$}
\put(190,80){$\Rightarrow N=U^TV =\left[\begin{smallmatrix}3 & 0 & 3\\1 & 3 & 0\end{smallmatrix}\right]$ 
}
\put(275,80){\scriptsize
 $,\;\mathbf{1}  \varphi(N)  \mathbf{1}^T =9$
}
\put(218,65){\scriptsize
$N \mathbf{1}^T = [6,4]$
}
\put(275,65){\scriptsize
$,\;\mathbf{1}  \varphi(N  \mathbf{1}^T )= 21$
}
\put(224,55){\scriptsize
$\mathbf{1} N =[4,3,3]$
}
\put(275,55){\scriptsize
 $,\;\varphi(\mathbf{1} N) \mathbf{1}^T = 12$
}
\put(210,40){\scriptsize
$\Rightarrow \GAM = 0.667 ,\; \AG=0.312$
}
\put(25,25){$VV^T$}
\put(0,0){\tiny 
$\left[\begin{smallmatrix}1 & 1 & 1 & 1 & 0 & 0 & 0 & 0 & 0 & 0\\1 & 1 & 1 & 1 & 0 & 0 & 0 & 0 & 0 & 0\\1 & 1 & 1 & 1 & 0 & 0 & 0 & 0 & 0 & 0\\1 & 1 & 1 & 1 & 0 & 0 & 0 & 0 & 0 & 0\\0 & 0 & 0 & 0 & 1 & 1 & 1 & 0 & 0 & 0\\0 & 0 & 0 & 0 & 1 & 1 & 1 & 0 & 0 & 0\\0 & 0 & 0 & 0 & 1 & 1 & 1 & 0 & 0 & 0\\0 & 0 & 0 & 0 & 0 & 0 & 0 & 1 & 1 & 1\\0 & 0 & 0 & 0 & 0 & 0 & 0 & 1 & 1 & 1\\0 & 0 & 0 & 0 & 0 & 0 & 0 & 1 & 1 & 1\end{smallmatrix}\right]$
}
\put(60,25){$-$}
\put(60,0){$-$}
\put(91,25){$UU^T$}
\put(66,0){\tiny $\left[\begin{smallmatrix}1 & 1 & 1 & 0 & 0 & 0 & 0 & 1 & 1 & 1\\1 & 1 & 1 & 0 & 0 & 0 & 0 & 1 & 1 & 1\\1 & 1 & 1 & 0 & 0 & 0 & 0 & 1 & 1 & 1\\0 & 0 & 0 & 1 & 1 & 1 & 1 & 0 & 0 & 0\\0 & 0 & 0 & 1 & 1 & 1 & 1 & 0 & 0 & 0\\0 & 0 & 0 & 1 & 1 & 1 & 1 & 0 & 0 & 0\\0 & 0 & 0 & 1 & 1 & 1 & 1 & 0 & 0 & 0\\1 & 1 & 1 & 0 & 0 & 0 & 0 & 1 & 1 & 1\\1 & 1 & 1 & 0 & 0 & 0 & 0 & 1 & 1 & 1\\1 & 1 & 1 & 0 & 0 & 0 & 0 & 1 & 1 & 1\end{smallmatrix}\right]$}
\put(160,25){$\Delta $}
\put(127,25){$=$}
\put(127,0){$=$}
\put(135,0){\tiny$ 
\left[\scriptsize \begin{smallmatrix}0 & 0 & 0 & 1 & 0 & 0 & 0 & 1 & 1 & 1\\0 & 0 & 0 & 1 & 0 & 0 & 0 & 1 & 1 & 1\\0 & 0 & 0 & 1 & 0 & 0 & 0 & 1 & 1 & 1\\1 & 1 & 1 & 0 & 1 & 1 & 1 & 0 & 0 & 0\\0 & 0 & 0 & 1 & 0 & 0 & 0 & 0 & 0 & 0\\0 & 0 & 0 & 1 & 0 & 0 & 0 & 0 & 0 & 0\\0 & 0 & 0 & 1 & 0 & 0 & 0 & 0 & 0 & 0\\1 & 1 & 1 & 0 & 0 & 0 & 0 & 0 & 0 & 0\\1 & 1 & 1 & 0 & 0 & 0 & 0 & 0 & 0 & 0\\1 & 1 & 1 & 0 & 0 & 0 & 0 & 0 & 0 & 0\end{smallmatrix}\right]$
}
\put(210,20){\scriptsize
$\Rightarrow \|\Delta\|_F^2 = 30$
}
\put(230,10){\scriptsize
$ NF_{RI} = 90 \Rightarrow \GAM=0.667$
}
\put(220,0){\scriptsize
${\|{VV^T}'\|}_F^2=|{VV^T}'|= 24$
}
\put(220,-10){\scriptsize
$\|{UU^T}'\|_F^2=|{UU^T}'|= 42$
}
\put(230,-20){\scriptsize
$ NF_{ARI} = 43.5 \Rightarrow \AG=0.312$
}
\end{picture}
\vspace{25pt}
\caption{\footnotesize Example for contingency v.s. co-membership based formulation. The $(A)RI$ is first derived from the contingency table $N$, using $\GAM$ formula where $\varphi(x)= x(x-1)/2$. Then same results are derived from the comparison of co-membership matrices $UU^T$ and $VV^T$, using the alternative formulation of $\GAM$, where $A_{n\times n}' = A-\mathbf{I}_{n}$ (see \reffoot{foot:exactRI} for details).
}
\label{fig:exDis}
\vspace{-10pt}
 \end{figure}
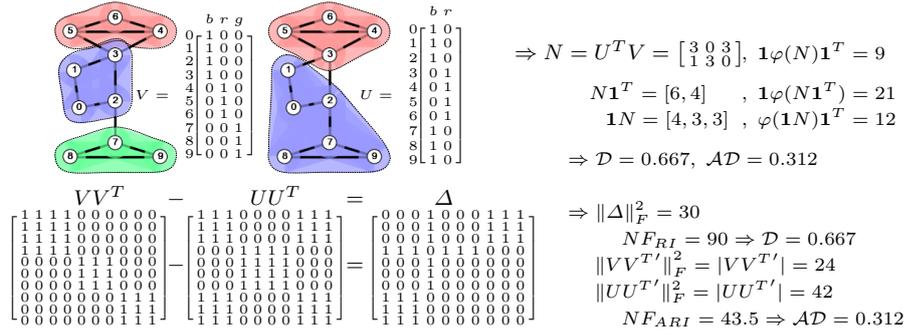
These formulations based contingency matrix of $U^TV$, as discussed in \refsec{sec:overlappSur}, are only appropriate for disjoint clusters. Therefore we propose the following reformulation of \refdef{def:deltaARI},  which is valid for both disjoint and overlapping cases. Instead of overlap matrix $U^TV$,  \refdef{def:deltaARI} measures the distance between clusterings directly from the difference of their co-membership matrices, i.e. $UU^T-VV^T$. This is inspired by the analogy between co-membership and overlap, i.e. $(UU^T)_{ij}$ denotes in how many clusters node $i$ and $j$ appeared together, and $(U^TU)_{ij}$ denotes how many nodes clusters $i$ and $j$ have in common.

\begin{definition}
\label{def:deltaARI} Co-Membership Clustering Difference ($\Delta$)
\[\Delta(U,V) = UU^T - VV^T,  \quad  \delta(U,V)= \frac{\Phi(\Delta)}{NF(U,V)}\]
where $\Phi: \mathbb{R}^{n\times n} \rightarrow \mathbb{R}$ is a matrix function which quantifies the qiven difference matrix, e.g. a matrix norm, and $NF(U,V)$ is a normalizing factor or an upper bound for $\Phi(U,V)$.
\end{definition}
\begin{theorem} \label{th:deltaeqRI}
For disjoint clusters, the approximate\footnote{\label{foot:exactRI}The exact formula derive if we change $n^2$ by $n(n-1)$ for the $RI$, and for the $ARI$ (\refeqb{eq:ari}) to also set the diagonal elements of the co-membership matrices to zero, i.e. ${UU^T}' = UU^T - \mathbf{I}_{n}$. 
Since the original $(A)RI$ formula counts only the co-memberships of pairs of nodes -- $(i,j)$  where $i\neq j$. The approximate version also considers the co-memberships for each single node with itself in different clusters, which is more suitable for overlapping cases.} RI and ARI (\refeqb{eq:ariapprox}) derive form $\Delta$ (proof in \refap{app:RIARI}), i.e. 
\begin{align*}
  & \Phi=  \| . \|^2_F \wedge \;   NF=n^2\times max(max(UU^T),max(VV^T))  \; \Rightarrow \; 1 - \delta \equiv  RI'(U,V) \\
  & \Phi=  \| . \|^2_F \wedge \;  NF=\|UU^T\|^2_F+\|VV^T\|^2_F-2 \frac{|UU^T||VV^T| }{n^2}\; \Rightarrow \;  1 - \delta  \equiv ARI'(U,V)  
\end{align*}
 where $|.|$ is sum of all elements in the matrix, and $\|.\|^2_F$ is the sum of squared values, a.k.a. squared Frobenius norm.
\end{theorem}

\begin{figure}
\begin{subfigure}[b]{0.45\textwidth}
\begin{picture}(80,50)
\put(0,0){\includegraphics[width=.95\textwidth]{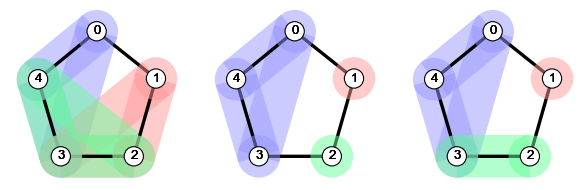}}
\put(0,42){$V$}\put(49,42){$U_1$}\put(95,42){$U_2$}
\put(0,-20){\scriptsize
 \setlength{\tabcolsep}{2pt}
 {\begin{tabular}{c|cccccc}
&$\omega$ & $A\omega$ & $RI_{\delta}$ &   $ARI_{\delta}$ &  $RI'_{\delta}$ &   $ARI'_{\delta}$ \\
\hline
$(V, U_1)$ & 0.5 & 0.22 &   0.80 &  0.25 &  0.90 &   0.32\\
$(V,U_2)$  & 0.5 &  0.19 &   0.88 &  0.49 &  0.94 &   0.58  \\
\end{tabular}}}
\end{picture}\vspace{30pt}
\caption{\footnotesize
Revisit to \reffig{fig:omegaexm}. 
Reported in the table are values for Omega index ($\omega$), and its adjusted version ($A\omega$), followed by our exact and approximate (marked by $'$) $\delta$-based $(A)RI$, derived from the proposed clustering co-memberships distance $\Delta$. 
}
\label{fig:overRevistitedOmegA}
\end{subfigure}\hspace{10pt}
\begin{subfigure}[b]{0.5\textwidth}
\begin{picture}(90,50)
\put(0,0){\includegraphics[width=\textwidth]{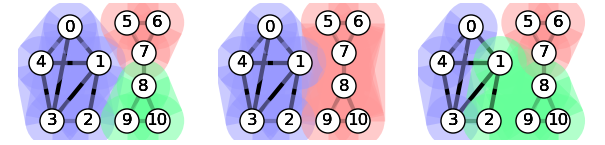}}
\put(0,42){$V$}\put(60,42){$U_1$}\put(115,42){$U_2$}
\put(-10,-20){\hspace{5pt}\scriptsize
 \setlength{\tabcolsep}{2pt}
 {\begin{tabular}{c|ccccccc}
 &{\tiny$NMI$} &{\tiny $NMI'$} & {\tiny $NMI''$} &  $RI_{\delta}$ &   $ARI_{\delta}$ &  $RI'_{\delta}$ &    $ARI'_{\delta}$\\
\hline
 $(V, U_1)$&0.78 & 0.61 & 0.53 &0.84 & 0.66 & 0.85 & 0.70 \\
$(V, U_2)$& 0.71 &  0.62& 0.61 &0.78 & 0.47& 0.80 & 0.57 \\
\end{tabular}}
}
\end{picture}\vspace{30pt}
\caption{\footnotesize
Revisit to \reffig{fig:setmatchingexm}. 
Here our exact and approximate co-membership based formulations are in agreement with the original non-overlapping $NMI$, and give a higher similarity score to $U_1$. Whereas the two overlapping $NMI$ extensions state the opposite.
}
\label{fig:overRevistitedmATCH}
\end{subfigure}
\caption{\footnotesize
Revisit to the examples of \reffig{fig:overlappingExamples}. 
On the left we see that Omega index ($\omega$) is unable to differentiate between $U_1$ and $U_2$, whereas its adjusted version even gives higher score to $U_1$, which is the opposite of what we expect. The fact that $U_2$ is more similar to $V$ is captured by our $\delta$-based $(A)RI$. 
On the right we see an example of disagreement between the original $NMI$ and its two set-matching based extensions for overlapping cases. 
Here since the problem is disjoint, $(A)RI_\delta$ gives same results as the original $(A)RI$. 
}
\label{fig:overRevistited}
 \vspace{-12pt}
\end{figure}

Our $\delta$-based formulation for $RI$ and $ARI$, presented in \refthe{th:deltaeqRI}, are also valid for overlapping cases.
These formulations denoted respectively by $RI_\delta$ and $ARI_\delta$ hereafter, are identical to the original formulations if clusterings are disjoint; whereas unlike the overlap based formulations, they always return $1$ if the clusterings are identical, regardless of the amount of the overlapping nodes. 
 Refer to \refap{app:RIARI} for more details, and see \reffig{fig:overRevistited} for examples.

It is worth mentioning that for crisp overlapping clusters, the Omega Index($\omega$) \cite{collins1988omega} derives from our formulation if we define $\Delta = [UU^T == VV^T]$, 
 i.e. $\Delta_{ij} =1$ if $(UU^T)_{ij} == (VV^T)_{ij}$ and zero otherwise. Then 
 \vspace{-5pt}
\begin{equation*}\scriptsize
\omega = |\Delta| - tr(\Delta),\quad A(\omega)=\frac{\omega-E[\omega]}{1-E[\omega]},\quad  E[\omega]= \sum_{i=0}^{min(r,k)} f_{UU^T}(i)f_{VV^T}(i)
\vspace{-5pt}
\end{equation*}
where $f_A(i)$ is the frequency of $i$ in $A$. 
\reffig{fig:overRevistitedOmegA} illustrates the effect of ignoring partial agreements by the $\omega$ index. 
%
Similarly, we can compute other normalized forms of $\Delta$, or compare the co-membership matrices of $UU^T$ and $VV^T$ in other ways, e.g using matrix divergences \cite{dhillon2007matrix,kulis2009low}. 
%
Here, we consider these two variations:
\begin{equation*}\scalebox{1}{$
\mathcal{D}_{norm} = \frac{ \|UU^T-VV^T\|^2_F}{\|UU^T\|^2_F+\|VV^T\|^2_F }\,,\quad
 I_{\sqrt{tr}} = \frac{tr(UU^TVV^T) }{\sqrt{tr((UU^T)^2)tr((VV^T)^2)}}= \frac{|UU^T \circ VV^T|  }{\|UU^T\|^2_F \|VV^T\|^2_F}$}
\end{equation*} 
It is also worth pointing out that in some applications, such as ensemble or multi-view clustering, we may not need the normalization and a measure of distance may suffice.
\vspace{-8pt}
\subsection{Extension for Network Clustering}\vspace{-5pt}
\label{sec:AlgextNet}
\vspace{-3pt}
Here we define structure dependent clustering distances which incorporate the underlying structure of the graph.
Let $N$ denote the incidence matrix of the graph $G$, such that $N_{ik} = \sqrt{A_{ij}}$ if node $i$ is incident with edge $k=(i,j)$, and zero otherwise.
Assuming a clustering as a transformation which assigns each datapoint to one of its $k$ clusters, i.e. $U: n\mapsto k$. 
, we can incorporate the structure by measuring the distance between the transformed data by $U$ and $V$ as:
\vspace{-5pt}
\begin{equation*}\GAM_\bot(U,V| G) = \GAM(N^TU, N^TV)\end{equation*} 
This is similar to measuring the structure similarity by counting the edges of the subgraphs, proposed earlier in \refsec{sec:intergam};  See \reffig{fig:trans} for an example. We should note that the above formulation requires an overlapping distance, such as $ARI_\delta$.
\begin{figure}
\begin{picture}(250,140)
\put(27,0){\includegraphics[width=.8\textwidth]{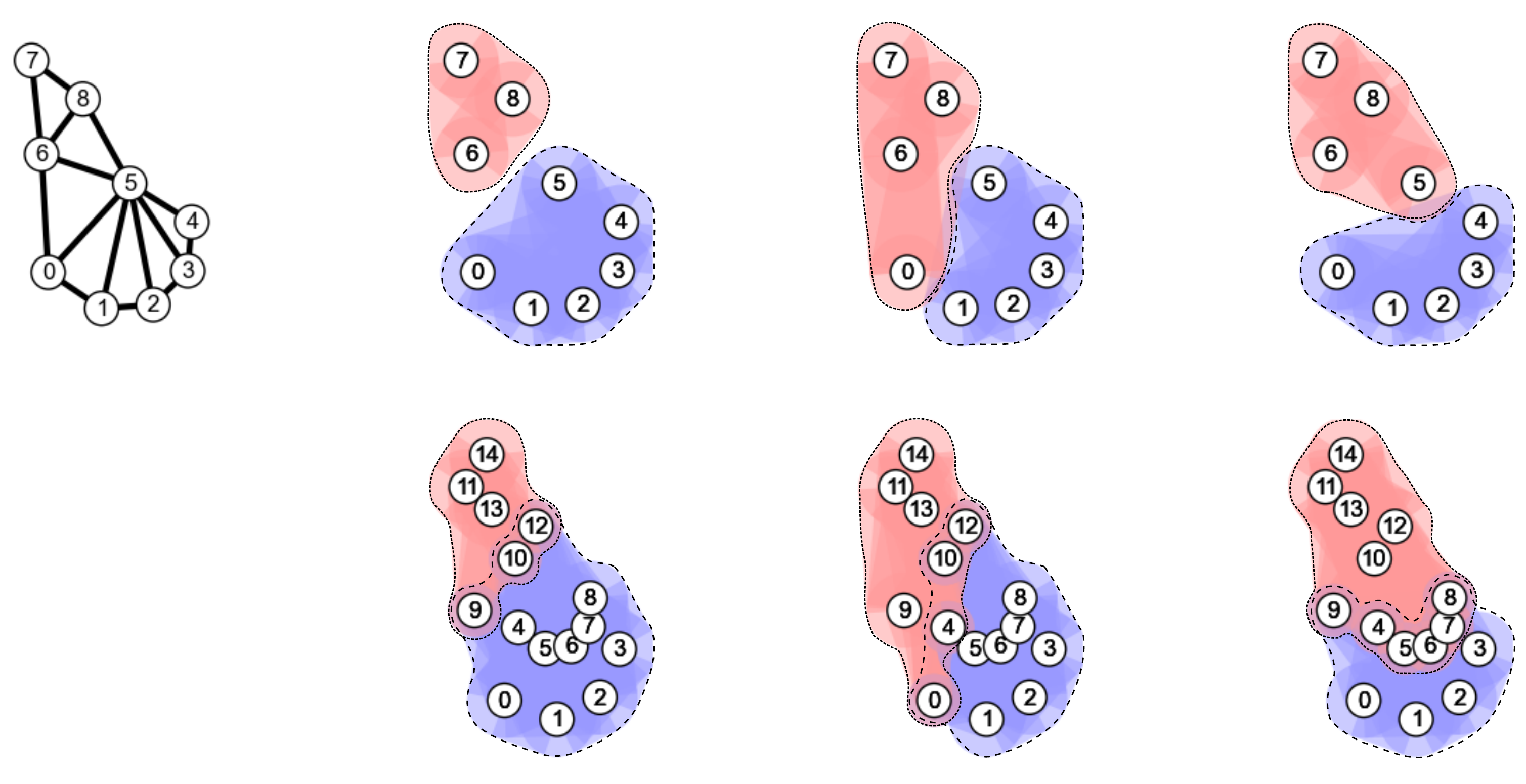}}
\put(15,120){\scriptsize$G:$}
\put(5,42){\tiny$N^T=\left[\begin{smallmatrix}1  & 1  & 0  & 0  & 0  & 0  & 0  & 0  & 0 \\0  & 1  & 1  & 0  & 0  & 0  & 0  & 0  & 0 \\0  & 0  & 1  & 1  & 0  & 0  & 0  & 0  & 0 \\0  & 0  & 0  & 1  & 1  & 0  & 0  & 0  & 0 \\1  & 0  & 0  & 0  & 0  & 1  & 0  & 0  & 0 \\0  & 1  & 0  & 0  & 0  & 1  & 0  & 0  & 0 \\0  & 0  & 1  & 0  & 0  & 1  & 0  & 0  & 0 \\0  & 0  & 0  & 1  & 0  & 1  & 0  & 0  & 0 \\0  & 0  & 0  & 0  & 1  & 1  & 0  & 0  & 0 \\1  & 0  & 0  & 0  & 0  & 0  & 1  & 0  & 0 \\0  & 0  & 0  & 0  & 0  & 1  & 1  & 0  & 0 \\0  & 0  & 0  & 0  & 0  & 0  & 1  & 1  & 0 \\0  & 0  & 0  & 0  & 0  & 1  & 0  & 0  & 1 \\0  & 0  & 0  & 0  & 0  & 0  & 1  & 0  & 1 \\0  & 0  & 0  & 0  & 0  & 0  & 0  & 1  & 1 \end{smallmatrix}\right]$
}
%
\put(137,113){\tiny $V=\bbordermatrix{~&b&r\cr 0& 1 & 0\cr 1 &1 & 0\cr 2 &1 & 0\cr 3  &1 & 0\cr 4 &1 & 0\cr 5 &1 & 0\cr 6 &0 & 1\cr 7 &0 & 1\cr 8 &0 & 1}$}
\put(217,113){\tiny $U_1=\bbordermatrix{~&b&r\cr 0& 0 & 1\cr 1 &1 & 0\cr 2 &1 & 0\cr 3 &1 & 0\cr 4 &1 & 0\cr 5 &1 & 0\cr 6 &0 & 1\cr  7 &0 & 1\cr 8 &0 & 1}$}
\put(295,113){\tiny $U_2=\bbordermatrix{~&b&r\cr 0 & 1 & 0\cr 1 &1 & 0\cr 2 &1 & 0\cr 3 &1 & 0\cr 4 &1 & 0\cr 5 &0 & 1\cr 6 &0 & 1\cr 7 &0 & 1\cr 8 &0 & 1}$}
\put(130,53){\tiny$N^TV=$}
\put(152,33){\tiny$\bbordermatrix{~&b&r\cr 0 & 2  & 0 \cr 1 &2  & 0 \cr 2 &2  & 0 \cr 3 &2  & 0 \cr 4 &2  & 0 \cr 5 &2  & 0 \cr 6 &2  & 0 \cr 7 &2  & 0 \cr 8 &2  & 0 \cr 9 &1  & 1 \cr 10 &1  & 1 \cr 11 &0  & 2 \cr  12 &1  & 1 \cr 13 &0  & 2 \cr 14 &0  & 2 }$
}\put(207,53){\tiny$N^TU_1=$}
\put(233,33){\tiny$\bbordermatrix{~&b&r\cr 0 & 1  & 1 \cr 1 &2  & 0 \cr 2 &2  & 0 \cr 3 &2  & 0 \cr 4 &1  & 1 \cr 5 &2  & 0 \cr 6 &2  & 0 \cr 7 &2  & 0 \cr 8 &2  & 0 \cr 9 &0  & 2 \cr 10 &1  & 1 \cr 11 &0  & 2 \cr 12 &1  & 1 \cr 13 &0  & 2 \cr 14 &0  & 2 }$}
\put(288,53){\tiny $N^TU_2=$}
\put(315,33){\tiny $\bbordermatrix{~&b&r\cr 0 &2  & 0  \cr 1 & 2   & 0  \cr 2& 2   & 0  \cr 3 & 2   & 0  \cr 4 & 1   & 1  \cr  5 & 1   & 1  \cr 6 & 1   & 1  \cr 7 & 1   & 1  \cr 8 & 1   & 1  \cr 9 & 1   & 1  \cr 10 & 0   & 2  \cr  11& 0   & 2  \cr 12 & 0   & 2  \cr 13 & 0   & 2  \cr 14 & 0   & 2  }$
}\vspace{5pt}
\end{picture}
\caption{\footnotesize A revisits to the example of \reffig{fig:exm}. Top) In the original data and considering only nodes, $U_1$ and $U_2$ have the same agreement with $V$. 
Since both $U_1$ and $U_2$ have one node clustered differently than $V$. 
Bottom) Transformed data using corresponding clusterings correctly identifies that $U_1$ is closer to $V$ compared to $U_2$. Note that the transformed data is similar to the line graph (edges as nodes) of the original data. 
}
\label{fig:trans}
\vspace{-2pt}
\end{figure}
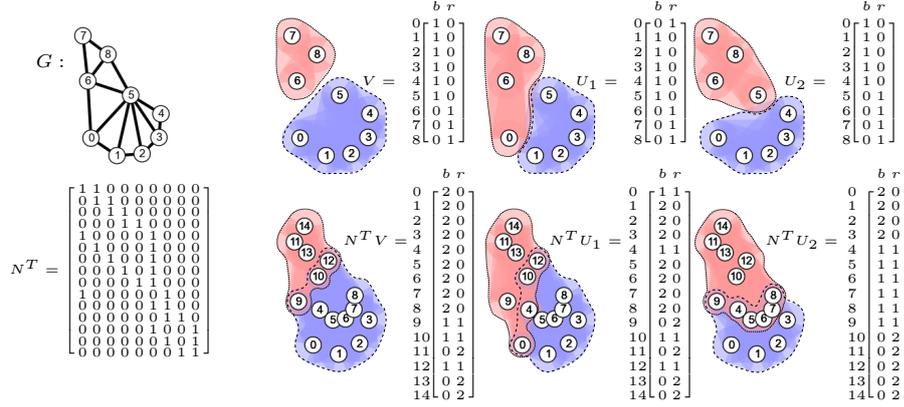
\begin{table}
\centering
\begin{tabular}{cccccccH}
  -        &    $RI_\delta$ &   $ARI_\delta$ &   $RI'_\delta$ &   $ARI'_\delta$ &    $I_{norm}$ &   $I_{\sqrt{tr}}$ &    $I_{tr}$ \\
\hline
\hline
 $(U,V_1)$     & 0.778 & 0.556 & 0.802 &  0.604 & 0.695 &  0.815 & 0.432 \\
 $(U,V_2)$    & 0.778 & 0.556 & 0.802 &  0.604 & 0.695 &  0.815 & 0.432 \\
 \hline
 $_\bot(U,V_1|G)$  & 0.926 & 0.744 & 0.928 &  0.752 & 0.799 &  0.923 & 0.527 \\
 $_\bot(U,V_2|G)$ & 0.857 & 0.417 & 0.859 &  0.435 & 0.708 &  0.844 & 0.480 \\
 $_{+}(U,V_1|G)$ & 0.889 & 0.773 & 0.901 &  0.797 & 0.843 &  0.904 & 0.712 \\
 $_{+}(U,V_2|G)$ & 0.833 & 0.660 & 0.900 &  0.776 & 0.832 &  0.885 & 0.705 \\
\hline
 $(N,U)$     & 0.750 & 0.500 & 0.979 &  0.327 & 0.512 &  0.662 & 0.200 \\
 $(N,V_1)$     & 0.750 & 0.491 & 0.979 &  0.337 & 0.503 &  0.668 & 0.193 \\
 $(N,V_2)$    & 0.639 & 0.264 & 0.977 &  0.275 & 0.481 &  0.616 & 0.178 \\
\hline
\end{tabular}
\caption{\footnotesize Results of different agreements for the example of \reffig{fig:exm}.   
The first two rows show that all the original structure independent measures result in the same agreement for $U_1$ and $U_2$. 
Whereas the structure based measures give higher agreement score to $U_1$  compared to $U_2$. The last three rows give the agreement of each clustering with the structure of the graph. 
}\label{tab:graphexm}
\vspace{-5pt}
\end{table}

Alternatively, we can assume each edge as a cluster of two nodes, and measure the distance of a clustering from the underlying structure of the graph. 
%
%
%
Consequently, the structure dependent distance of $U$ and $V$ can be defined as a combination of $\GAM(U, N)$, $\GAM(V, N)$ and $\GAM(U, V)$, 
for example:
\vspace{-5pt}\[\GAM_+(U,V| G) = \alpha\GAM(U, V)+(1-\alpha)|\GAM(U, N)-\GAM(V, N)|, \quad \alpha = 0.5\]
%
%

\reftab{tab:graphexm}, \reftab{tab:graphexmOmega} and \reftab{tab:graphexmMatch} compare structure dependent and independent measures for our earlier examples in \reffig{fig:exm}, and  \reffig{fig:overlappingExamples}. 
Wherein the experiments of the next section compare the measures in the context of community mining evaluation. 

\begin{table}
\centering
\begin{tabular}{cccccccH}
  -        &    $RI_\delta$ &   $ARI_\delta$ &   $RI'_\delta$ &   $ARI'_\delta$ &    $I_{norm}$ &   $I_{\sqrt{tr}}$ &    $I_{tr}$ \\
\hline\hline
 $(V,U_1)$     & 0.800 & 0.245 & 0.902 &  0.318 & 0.532 &  0.764 & 0.378 \\
 $(V,U_2)$    & 0.875 & 0.490 & 0.942 &  0.577 & 0.663 &  0.894 & 0.444 \\
\hline
 $_\bot(V,U_1|G)$  & 0.856 & 0.186 & 0.868 &  0.211 & 0.536 &  0.860 & 0.538 \\
 $_\bot(V,U_2|G)$ & 0.913 & 0.427 & 0.924 &  0.483 & 0.672 &  0.961 & 0.581 \\
 $_{+}(V,U_1|G)$  & 0.775 & 0.556 & 0.919 &  0.617 & 0.720 &  0.859 & 0.662 \\
 $_{+}(V,U_2|G)$ & 0.863 & 0.712 & 0.954 &  0.765 & 0.824 &  0.945 & 0.706 \\
\hline
 $(N,U)$     & 0.850 & 0.333 & 0.933 &  0.528 & 0.682 &  0.816 & 0.333 \\
 $(N,U_1)$     & 0.600 & 0.200 & 0.870 &  0.444 & 0.590 &  0.771 & 0.280 \\
 $(N,U_2)$    & 0.700 & 0.400 & 0.900 &  0.576 & 0.666 &  0.822 & 0.300 \\
\hline
\end{tabular}
\caption{\footnotesize Results of different agreements for the omega example of \reffig{fig:omegaexm}. 
}\label{tab:graphexmOmega}
\end{table}
\begin{table}
\centering
\begin{tabular}{cccccccH}
  -        &    $RI_\delta$ &   $ARI_\delta$ &   $RI'_\delta$ &   $ARI'_\delta$  &    $I_{norm}$ &   $I_{\sqrt{tr}}$ &    $I_{tr}$ \\
\hline\hline
 $(V,U_1)$     & 0.836 & 0.660 & 0.851 &  0.703 & 0.705 &  0.840 & 0.355 \\
 $(V,U_2)$    & 0.782 & 0.471 & 0.802 &  0.567 & 0.626 &  0.721 & 0.256 \\
 \hline
 $_\bot(V,U_1|G)$  & 0.900 & 0.790 & 0.906 &  0.806 & 0.768 &  0.902 & 0.407 \\
 $_\bot(V,U_2|G)$ & 0.857 & 0.564 & 0.862 &  0.607 & 0.667 &  0.798 & 0.305 \\
 $_{+}(V,U_1|G)$  & 0.855 & 0.708 & 0.922 &  0.793 & 0.839 &  0.866 & 0.675 \\
 $_{+}(V,U_2|G)$ & 0.818 & 0.556 & 0.897 &  0.716 & 0.782 &  0.804 & 0.616 \\
\hline
 $(N,U)$     & 0.945 & 0.865 & 0.977 &  0.620 & 0.615 &  0.814 & 0.176 \\
 $(N,U_1)$     & 0.818 & 0.621 & 0.970 &  0.502 & 0.589 &  0.707 & 0.182 \\
 $(N,U_2)$    & 0.800 & 0.506 & 0.968 &  0.485 & 0.552 &  0.702 & 0.152 \\
\hline
\end{tabular}
\caption{\footnotesize Results of different agreements for the matching example of \reffig{fig:setmatchingexm}.
}\label{tab:graphexmMatch}
\end{table}

\vspace{-15pt}
\section{Experimental Results}\vspace{-5pt}
Clustering agreement measures are often used in external evaluation of clustering algorithms, i.e. to compare their results with the known ground-truth in the benchmark datasets~\cite{Lancichinetti09Comparison}.
Here we perform similar sets of experiments, however the purpose is not to compare the general performance of community mining methods, but rather to show different comparisons/rankings we obtained using different agreement measures. 
Three sets of results are presented in the following to compare i) classic agreement indexes, ii) structure dependent and independent indexes, and iii) overlapping extensions.
 \vspace{-10pt}
 \subsection{Experiment Settings} \vspace{-5pt}
In each experiment we select a set of common community mining methods, which discover clusters in a given network from different methodologies. 
In case of disjoint partitioning for \refsec{sec:expClass} and \ref{sec:expStr}, we use Louvain by \citet{Vincent08Louvain}, WalkTrap by~\citet{pons2005computing}, PottsModel by \citet{ronhovde2009multiresolution}, FastModularity by \citet{newman2004fast}, and InfoMap by \citet{Rosvall08Infomap}. 
For \refsec{sec:expOver} we select four overlapping community detection methods: COPRA by \citet{gregory2010finding}, MOSES by \citet{mcdaid2010detecting}, 
OSLOM by \citet{lancichinetti2011finding}, and BIGCLAM by \citet{yang2013overlapping}. 
The authors' original implementations are used for all the algorithms, with no parameter tuning (defaults are used); and the reported agreements are averaged over ten runs. 

Datasets are generated using the LFR~\cite{Lancichinetti08LFR} benchmarks, which are commonly used in the evaluation of community mining algorithms. Parameters are chosen similar to the experiments by \citet{Lancichinetti09Comparison}, i.e. networks with 1000 nodes, average degree of 20, max degree of 50, and power law degree exponent of -2; where the size of communities follows a power law distribution with exponent of -1, and ranges between 20 to 100 nodes. 
For the first experiments in \refsec{sec:expClass}, we generated \emph{unweighted} LFR benchmarks with mixing parameters that varies from 0.1 to 0.8.  
Second experiment in \refsec{sec:expStr} uses \emph{weighted} LFR benchmarks~\cite{Lancichinetti09LFR}, where the mixing parameter for topology is fixed to $0.5$, and the mixing parameter for weights varies.
For the last experiment, we change the fraction of overlapping nodes, and generate unweighted LFR networks with the mixing parameter for topology fixed to $0.1$, and $2$ is set as the maximum number of communities a node can belong to, similar to experiments in \cite{lancichinetti2011finding}. Results for other parameter settings, including smaller sized communities (10 to 50), could be found in the supplementary materials\footnote{
\url{https://github.com/rabbanyk/CommunityEvaluation}}.
 \begin{figure}
 \vspace{-7pt}
\centering
\includegraphics[width=.9\textwidth]{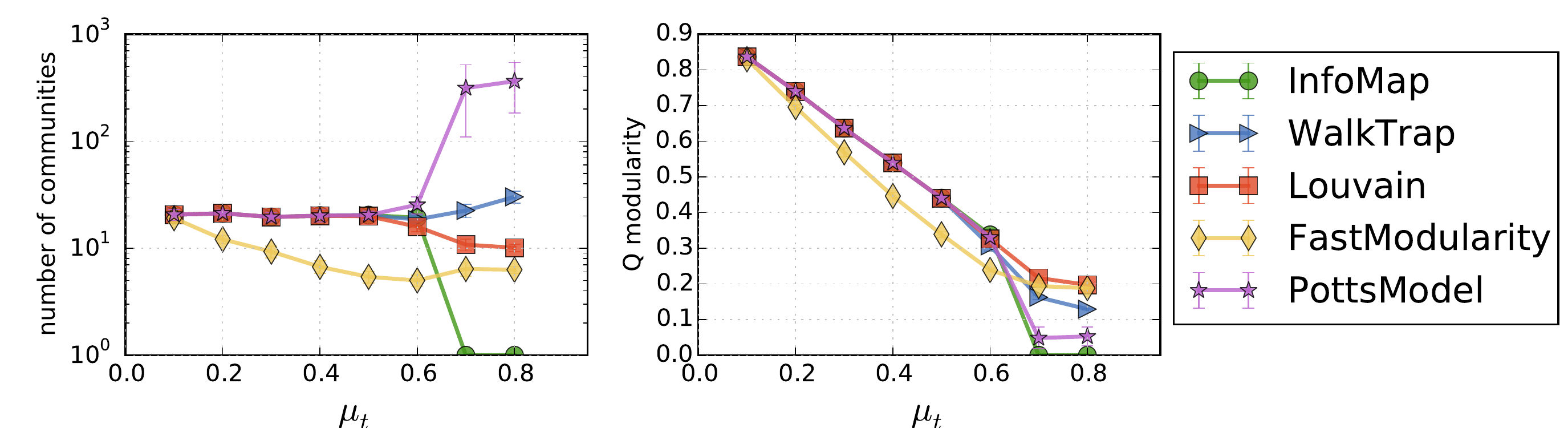}\vspace{-10pt}
\includegraphics[width=\textwidth]{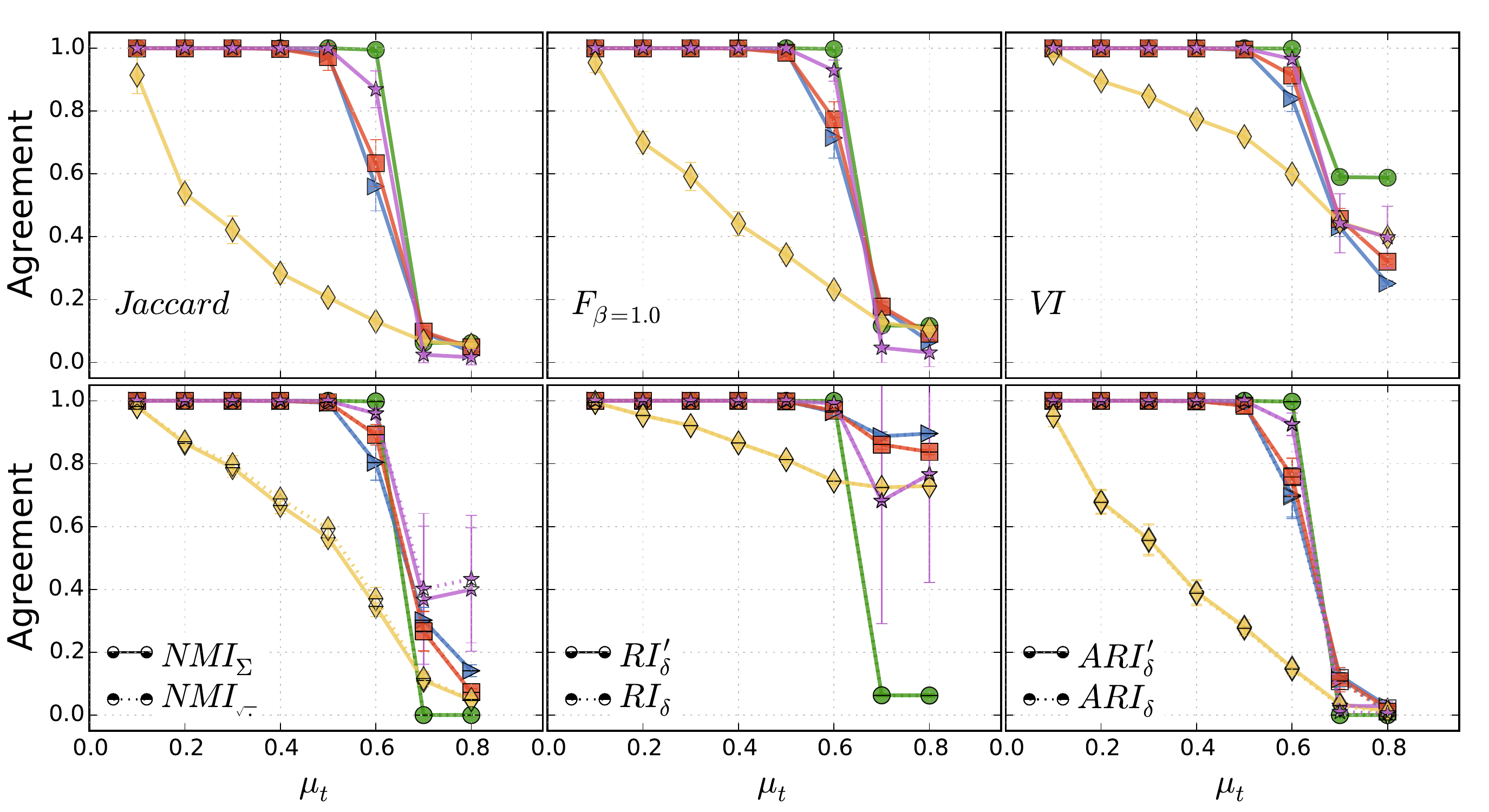}\vspace{-10pt}
\caption{
The agreement of results from different community detection algorithms with the ground-truth in \emph{unweighted LFR benchmarks}, plotted as a function of the mixing parameter.    
For large mixing parameters ($\mu_t$), $NMI_{\sqrt{}}$ and $NMI_{\Sigma}$ rank PottsModel significantly higher since it finds too many communities; whereas $VI$ (as opposite to $RI$) marks InfoMap significantly better mainly because it resulted in too few communities; neither are close to the ground-truth. 
In the last three plots, similar measures are overlaid to show they are highly similar.
}\vspace{-10pt}
\label{fig:uNMIARI}
\end{figure}
\vspace{-10pt}
 \subsection{Classic Measures}
 \vspace{-5pt}
 \label{sec:expClass}
~\reffig{fig:uNMIARI} shows the comparison of the algorithms obtained by six different agreement measures\footnote{Similar trends are observed for other variations of agreement measures which can be found in the supplementary materials.}.
 Overall, the ranking of the algorithms according to these agreement measures is very similar. 
 However, for large mixing parameters, the PottsModel is ranked significantly higher according to 
 the $NMI$ ($NMI_{\sum}$ or $NMI_{\sqrt{.}}$), which is not consistent with the ranking obtained from the $ARI$, plotted as $ARI_\delta$ in \reffig{fig:uNMIARI}. The $\delta$ subscript indicates that the $ARI$ is computed based on our $\delta$-based formulation, which is equivalent to the original $ARI$ in this experiment, since communities are non-overlapping (\refthe{th:deltaeqRI}). 
This disagreement for large mixing parameters is most probably because of the bias $NMI$ has to the larger number of clusters \cite{Vinh09AMI}.  
 Apart form this difference, the ranking from $NMI$ is very similar to the one obtained from $ARI$. This is expected as these indices are measuring the same quantity as shown in the generalization of \refdef{def:agam}. 
  We can further see that there is no clear difference between the rankings from the approximate (See \reffoot{foot:exactRI})
 and original $ARI$, i.e. $ARI_\delta'$ and $ARI_\delta$ in the \reffig{fig:uNMIARI}. This is desirable as we can use them interchangeably, whilst the former is more appropriate in the case of overlapping clusters, as discussed in~\refsec{sec:overlappAlgebra}.

\begin{figure}[h!]
\centering\vspace{-7pt}
\includegraphics[width=.9\textwidth]{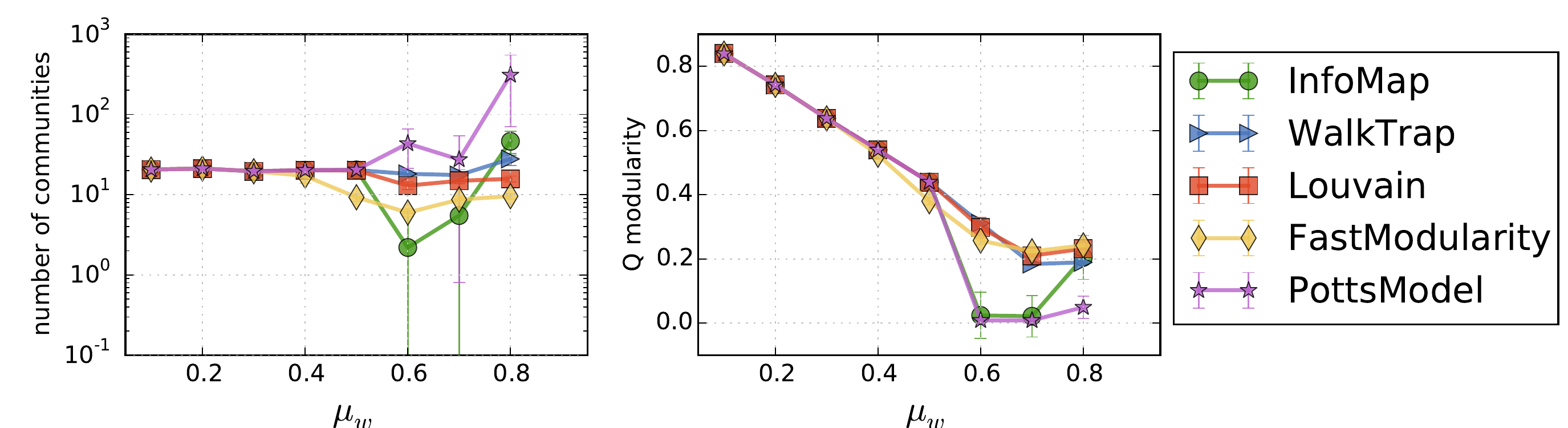}\vspace{-10pt}
\includegraphics[width=\textwidth]{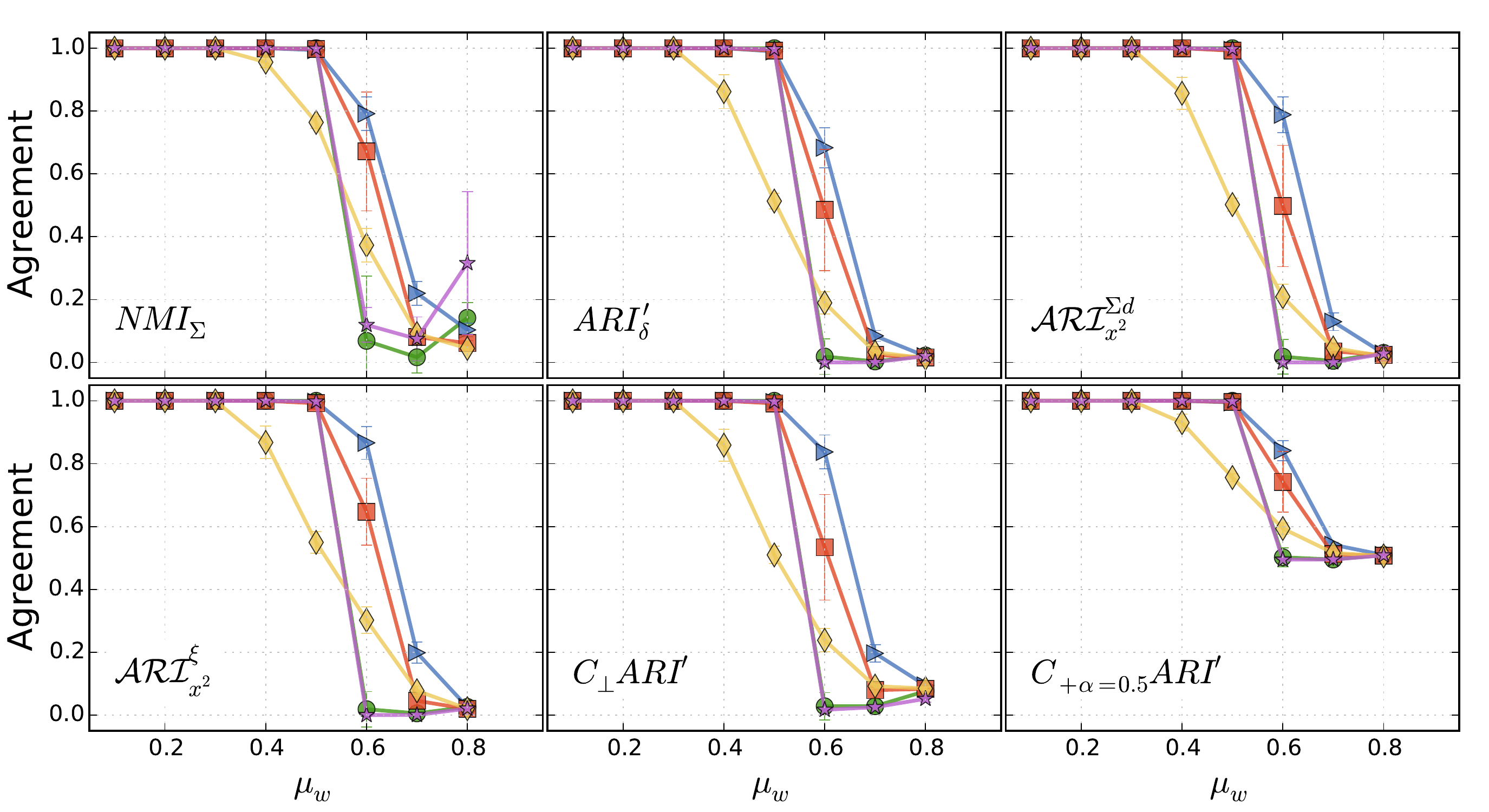}\vspace{-10pt}
\caption{Comparison of agreement indexes on \emph{weighted LFR benchmark}, plotted as a function of the mixing parameter for weights. The difference between WalkTrap and Louvain is more significant according to the $\mathcal{ARI}_{x^2}^{\xi} $, $\mathcal{ARI}_{x^2}^{\Sigma d}$, and $C_{\bot}ARI'$ which are structure dependent measures. We can also see the bias of $NMI$ to the number of clusters, similar to the \reffig{fig:uNMIARI}. }
\label{fig:wNMIARI}\vspace{-10pt}
\end{figure}

\vspace{-10pt}
 \subsection{Structure Dependent Measures}\vspace{-5pt}
 \label{sec:expStr}
\reffig{fig:wNMIARI} compares the community mining methods over the \emph{weighted} LFR benchmarks. 
Similar to the previous experiment, the rankings are very close. However, the difference between structure dependent and independents measures has become clear with the presence of weights. 
We can see that the Walktrap method is performing better according to most of the measures, whereas the distinction is more readable in the structure dependent variations: i.e. the degree weighted ARI ($\mathcal{ARI}_{x^2}^{\xi} $), and edge counting ARI ($\mathcal{ARI}_{x^2}^{\Sigma d}$) introduced in \refsec{sec:intergam}, and the transformed ARI ($C_{\bot}ARI'$) introduced in \refsec{sec:AlgextNet}. 

\begin{figure}[h!]\vspace{-10pt}
\centering
\includegraphics[width=.9\textwidth]{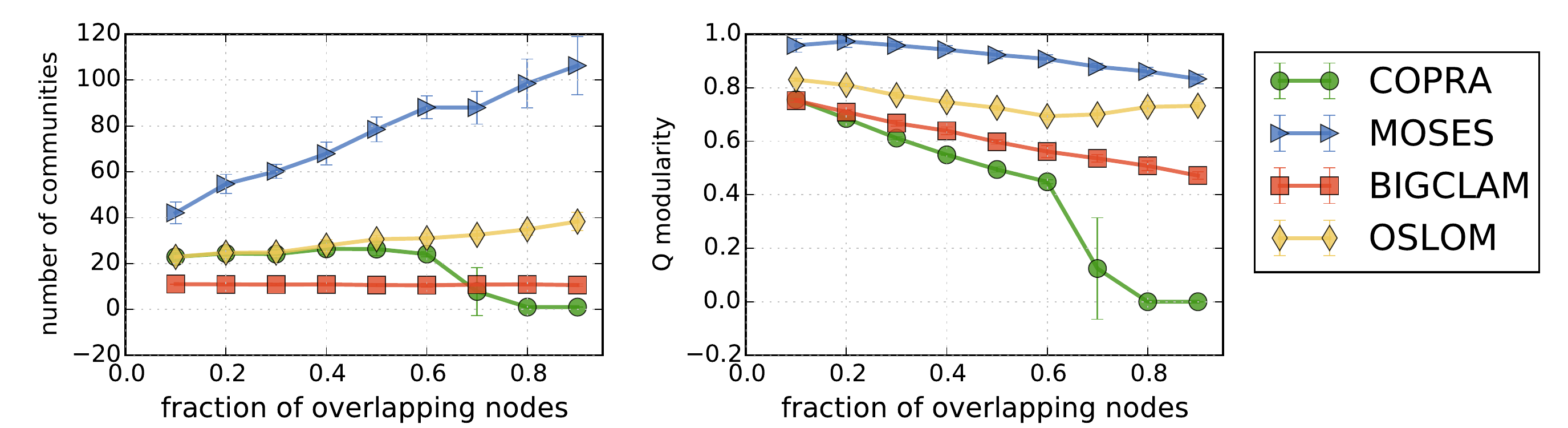}\vspace{-10pt}
\includegraphics[width=1\textwidth]{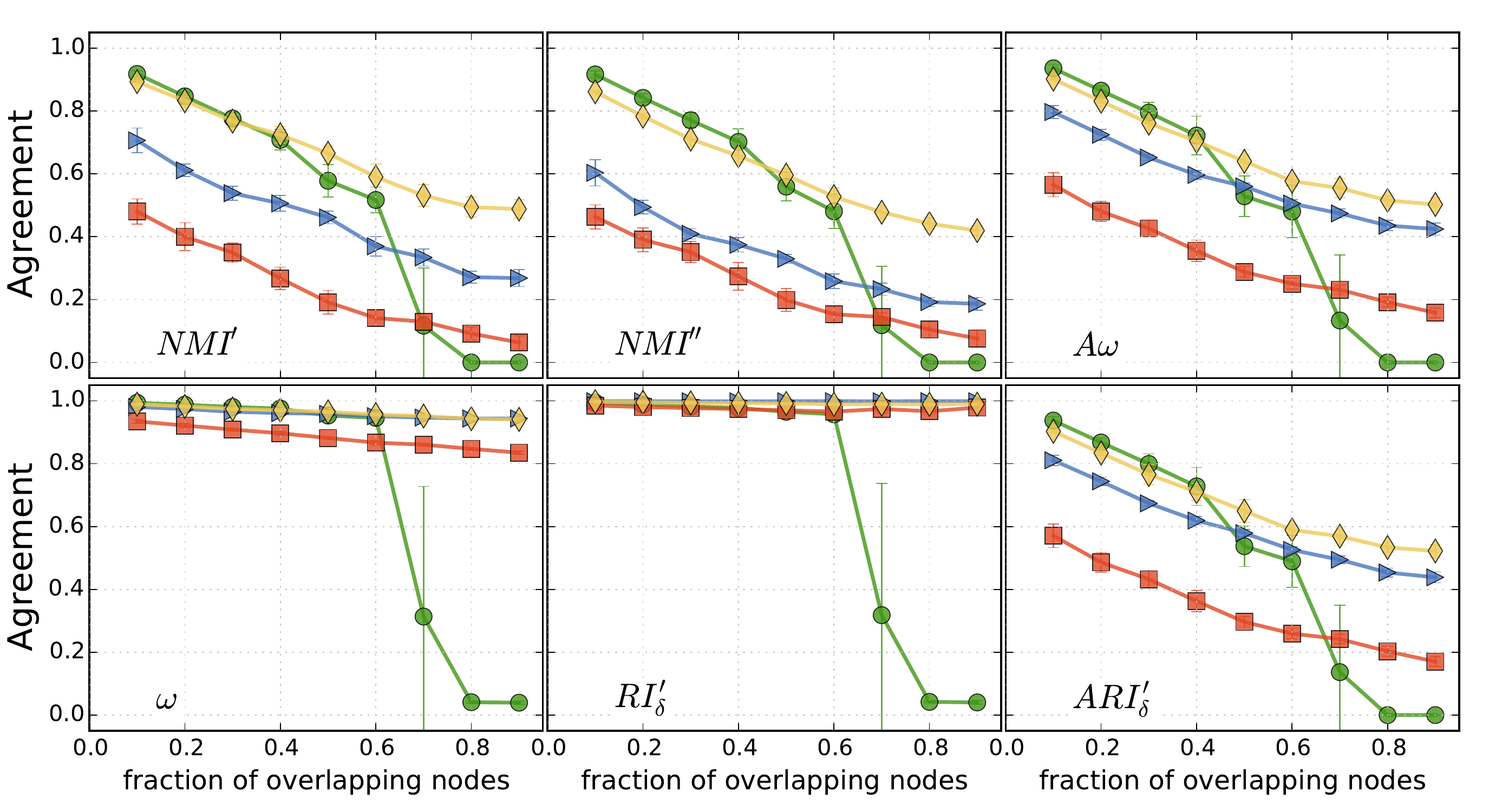}
\caption{Comparison of agreement indexes on \emph{overlapping LFR benchmark}, 
plotted as a function of the fraction of overlapping nodes. We can see the negative bias of number of clusters in the set matching overlapping measures, i.e. $NMI'$ and $NMI''$, which strongly penalize MOSES  for finding many communities. Which is not the case with $A\omega$ and $ARI'_\delta$. We can also see the impracticability of un-adjusted measures: $\omega$ and $RI_\delta'$, which are also very similar.  }
\label{fig:oNMIARI}
\vspace{-10pt}
\end{figure}
\vspace{-10pt}
 \subsection{Overlapping Measures}
 \label{sec:expOver}\vspace{-5pt}
\reffig{fig:oNMIARI} shows the comparison of these methods based on different overlapping agreement indexes: the overlapping NMI variations: $NMI'$ by \citet{Lancichinetti08overlap} and $NMI''$ from  \citet{mcdaid2011normalized}; the omega index ($\omega$), and its adjusted version ($A\omega$); and our $\delta$-based formulations for the $RI$ and $ARI$, i.e. $RI_\delta'$, and $ARI_\delta'$. 
Here also we observe a generally similar ranking. However the difference between MOSES and OSLOM is more significant according to the set-matching based extensions of $NMI$. This most probably is because MOSES finds much more communities, and hence it is more likely for it to have communities that do not get matched/compared with the communities in the ground-truth, although they show valid groupings of the nodes. We can also see that in this case, the ranking from adjusted omega, $A\omega$ and $ARI_\delta'$ are very similar, which can be explained as in our settings, each node can only belong to maximum of two communities; whereas the difference between $A\omega$ and $ARI_\delta'$ becomes clear if a node can belong to many communities.

\section{Conclusion}\vspace{-5pt}
In this paper, we presented generalizations of clustering agreement measures. This generalization illustrates the relation between the Rand Index ($RI$) and Variation of Information ($VI$); and Adjusted Rand Index ($ARI$) and the Normalized Mutual Information ($NMI$). We then discussed the necessity of structure dependent agreement measures, particularly in the evaluation of clusters over networks, i.e. communities; and proposed extensions of the general formula for such cases. We further discussed the difficulty of extending this contingency based formula for overlapping clusters and proposed reformulation which works for overlapping cases. We showed that the original $RI$ and $ARI$ of non-overlapping clusters derive from this reformulation.  
\bibliographystyle{spbasic}      
\bibliography{7_references}   

\appendix
\section{Proofs}
\label{app:proofs}
\vspace{-10pt}
\begin{minipage}{\textwidth}
\proofOf{Proposition \ref{th:varen}}
\label{app:proVIAI}
From the definition of Variation of information we have:\vspace{-5pt}
{\small
\[VI(U,V) =  H(U) + H(V) - 2I(U,V) = 2H(U,V)-H(U)-H(V) = \mathbf {H(V|U)+H(U|V)}\]}
\vspace{-3pt}On the other hand, we have:
{\scriptsize
\begin{align*}
 RI(U,V)
& \propto  \frac{1}{n^2-n}(\sum_{i=1}^k [\sum_{j=1}^r n_{ij}^2 - (\sum_{j=1}^r n_{ij})^2] + \sum_{j=1}^r [\sum_{i=1}^k n_{ij}^2 - (\sum_{i=1}^k n_{ij})^2])\\[-5pt]
&\overset{*}{\propto} \sum_{i=1}^k [E_j(n_{ij}^2) - E_j( n_{ij})^2] + \sum_{j=1}^r [E_i(n_{ij}^2) - E_i(n_{ij})^2]\\[-5pt]
&\overset{*}{\propto} \sum_{i=1}^k Var_j(n_{ij}) + \sum_{j=1}^r Var_i(n_{ij})
 \quad\overset{**}{\propto}  \quad\mathbf {  Var(V|U)+ Var(U|V) } \quad  \quad \qed
\end{align*}}
\end{minipage}
\begin{minipage}{\textwidth}
{\vspace{-5pt}
\scriptsize $(*)$ $E_j$/$Var_j$ shows the average/variance of values in the $j^{th}$ column of the contingency table.   \\
$(**)$ The RI is in fact proportional to the average variance of rows/columns values in the contingency table, which we denote by conditional variance. For other forms of conditional variance for categorical data see \citet{light71}.    
} 
\proofOf{\refco{co:norm}}
\label{app:norm}
We first show that $0 \leq \GAM_{\varphi}^\eta (U||V)$ which also results in the lower bound $0$ for $\GAM_{\varphi}^\eta(U,V)$ since, 
\( \GAM_{\varphi}^\eta(U,V) = \GAM_{\varphi}^\eta (U||V) + \GAM_{\varphi}^\eta (V||U) \).
From the superadditivity of $\varphi$ we have:\vspace{-5pt}
\begin{align*}
  & \sum_{u\in U} \varphi(\eta_{uv}) \leq  \varphi(\sum_{u\in U} \eta_{uv})  
    \implies 
    \sum_{v\in V} 
    \left[ \varphi(\sum_{u\in U} \eta_{uv})- \sum_{u\in U} \varphi(\eta_{uv}) \right] \geq 0 
   \implies  \mathbf {\GAM_{\varphi}^\eta (U||V) \geq 0}
\end{align*} 
Similarly for the upper bound, from positivity and super-additivity we get respectively:
$$\GAM_{\varphi}^\eta (U||V) = \sum_{v\in V}  \varphi(\sum_{u\in U} \eta_{uv})  -  \sum_{v\in V}  \sum_{u\in U} \varphi(\eta_{uv}) \leq \sum_{v\in V}  \varphi(\sum_{u\in U} \eta_{uv}) \leq  \varphi(\sum_{v\in V}  \sum_{u\in U} \eta_{uv}) \ $$ 
\proofOf{Identity \ref{id:vi}}
\label{app:idVI}
The proof is elementary, if we write the definition for $\varphi ={x}\log x$, we get:
\begin{align*}
 \NGAM_{x\log x}^{|\cap|}(U,V) & = \frac{\sum_{v\in V} \sum_{u\in U} |u\cap v|  \left[ \log(\sum_{u\in U} |u\cap v|) - \log(|u\cap v|) \right]}  {\left( \sum_{v\in V} \sum_{u\in U} |u\cap v| \right) \log   \left( \sum_{v\in V} \sum_{u\in U} |u\cap v| \right)}\\
& +  \frac{\sum_{u\in U} \sum_{v\in V} |u\cap v|
\left[\log(\sum_{v\in V} |u\cap v|) - \log(|u\cap v|) \right]}
 {\left( \sum_{u\in U} \sum_{v\in V}|u\cap v|\right) \log \left( \sum_{u\in U} \sum_{v\in V}|u\cap v|\right)}\\
  &   \overset{*}{=}  \frac{\sum_{j}^r \sum_{i}^k n_{ij} 
  \left[  \log(\sum_{i}^k n_{ij})+ \log(\sum_{j}^r n_{ij}) -2 \log(n_{ij})  \right]}
 {(\sum_{i}^k\sum_{j}^r n_{ij})\log (\sum_{i}^k\sum_{j}^r n_{ij})}\\
   & \overset{**}{=}  \frac{1}{\log n}\sum_{j}^r \sum_{i}^k \frac{ n_{ij} }{n} \log(\frac{ n_{i.} n_{.j}}{n_{ij}^2}) 
 =  \frac{VI(U,V)}{\log n}\quad  \quad \qed\vspace{-5pt}
\vspace{-5pt}\end{align*} 
{\scriptsize $(*)$ slight change of notation, i.e. from $\sum_{u\in U}$ to $\sum_{i}^k$, $\sum_{v\in V}$ to $\sum_{j}^r$ and $|u\cap v|$ to $n_{ij}$.\\
$(**)$ since $\sum_{i}^k\sum_{j}^r n_{ij} = n$, $\sum_{i}^k n_{ij} = n_{.j}$ and $\sum_{j}^r n_{ij} = n_{i.}$. } 
\proofOf{Identity \ref{id:ri}} Similar to the previous proof from the definition we derive:
\label{app:idRI}
\begin{align*}
 \NGAM_{\binom{x}{2}}^{|\cap|}(U,V) 
 & \overset{*}{=}   \frac{ \sum_{j}^r \left[ (\sum_{i}^k n_{ij})^2 -\sum_{i}^k n_{ij}^2 \right] + \sum_{i}^k \left[ (\sum_{j}^r n_{ij})^2 - \sum_{j}^r n_{ij}^2  \right]}
 {(\sum_{i}^k\sum_{j}^r n_{ij})^2-\sum_{i}^k\sum_{j}^r n_{ij}} \\[-3pt]
  &  \overset{**}{=}    \frac{1}{n^2-n} [ 
  \sum_{j}^r (n_{.j})^2 +  \sum_{i}^k (n_{i.})^2 - 2 \sum_{j}^r \sum_{i}^k n_{ij}^2  ]
   = 1-RI(U,V) \quad  \quad \qed
\vspace{-5pt}
\end{align*}
{\scriptsize $(*), (**)$ same as previous proof.  }
\end{minipage}
\noindent
\begin{minipage}{\textwidth}
\proofOf{Identity \ref{id:nmi} and \ref{id:ari}} 
\label{app:idNMI}
\begin{align*}
\label{eq:AGAMA}
&\nonumber\AGAM_{\varphi}^\eta = \frac{\sum_{v\in V} \varphi(\eta_{.v}) + \sum_{u\in U} \varphi(\eta_{u.}) -2 \sum_{v\in V} \sum_{u\in U} \varphi(\eta_{uv})}
{ \sum_{v\in V} \varphi(\eta_{.v}) + \sum_{u\in U} \varphi(\eta_{u.}) - 2 \sum_{u\in U}\sum_{v\in V} \varphi \left(\frac{\eta_{.v}\eta_{u.}}{\sum_{u\in U}\sum_{v\in V} \eta_{uv} }\right) }\\
\Rightarrow \; & 1 - \AG_{\varphi}^{\eta}(U,V) = \frac{  \sum_{v\in V} \sum_{u\in U} \varphi(\eta_{uv})-  \sum_{u\in U}\sum_{v\in V} \varphi \left(\frac{\eta_{.v}\eta_{u.}}{\sum_{u\in U}\sum_{v\in V} \eta_{uv} }\right)}
{ \frac{1}{2}\left[ \sum_{v\in V} \varphi(\eta_{.v}) + \sum_{u\in U} \varphi(\eta_{u.})\right] -  \sum_{u\in U}\sum_{v\in V} \varphi \left(\frac{\eta_{.v}\eta_{u.}}{\sum_{u\in U}\sum_{v\in V} \eta_{uv} }\right) }
\end{align*}
This formula resembles the adjuctment for chance in \refeqb{eq:adjust}, where the measure being adjusted is $\sum_{v\in V}\sum_{u\in U} \varphi(\eta_{uv})$, the upper bound used for it is $\frac{1}{2}[\sum_{v\in V} \varphi(\eta_{.v}) + \sum_{u\in U} \varphi(\eta_{u.})]$, and the expectation is defined as:
\[E[\sum_{v\in V}\sum_{u\in U} \varphi(\eta_{uv})] = \sum_{u\in U}\sum_{v\in V} \varphi \left(\frac{\eta_{.v}\eta_{u.}}{\sum_{u\in U}\sum_{v\in V} \eta_{uv} }\right) \]
Now if we have $\varphi(xy) = \varphi(x)\varphi(y)$, which is true for $\varphi(x)=x^2$, we have:
\[E[\sum_{v\in V}\sum_{u\in U} \varphi(\eta_{uv})] = \sum_{u\in U}\sum_{v\in V}  \frac{\varphi(\eta_{.v})\varphi(\eta_{u.})}{\varphi(\sum_{u\in U}\sum_{v\in V} \eta_{uv} )} =  \frac{\sum_{v\in V} \varphi(\eta_{.v})\sum_{u\in U}\varphi(\eta_{u.})}{\varphi(\sum_{u\in U}\sum_{v\in V} \eta_{uv} )} \]
Using this expecation, if we substitute $\varphi=x^2$ we get the $ARI'$ of \refeqb{eq:ariapprox}, and using the $\varphi=\binom{x}{2}$ and the later reformulation of $E$, we get the original $ARI$ of \refeqb{eq:ari}, as:
\begin{align*}
1 - \AG_{\binom{x}{2}}^{|\cap|}(U,V)& =\frac{	\sum\limits_{v\in V} \sum\limits_{u\in U} \binom{|u\cap v|}{2}
		- E(\sum\limits_{v\in V} \sum\limits_{u\in U} \binom{|u\cap v|}{2} )
}
{ \frac{1}{2}\left[
			\sum\limits_{v\in V}\binom{\sum\limits_{u\in U} |u\cap v|}{2}
		  + \sum\limits_{u\in U}\binom{\sum\limits_{v\in V} |u\cap v|}{2}
		\right] - E(\sum\limits_{v\in V} \sum\limits_{u\in U} \binom{|u\cap v|}{2})
}\\
& where \quad 
E(\sum\limits_{v\in V} \sum\limits_{u\in U} \binom{|u\cap v|}{2} )= 
\frac{
	\sum\limits_{v\in V}\binom{\sum\limits_{u\in U} |u\cap v|}{2}
		   \sum\limits_{u\in U}\binom{\sum\limits_{v\in V} |u\cap v|}{2}
	} 	{	
	\binom{ n}{2}
}\\
\Rightarrow 1 - \AG_{\binom{x}{2}}^{|\cap|}(U,V)& \overset{*,**}{=}
\frac{	\sum_j^r \sum_i^k \binom{n_{ij}}{2}
		- \sum_j^r \binom{n_{.j}}{2} \sum_i^k\binom{n_{i.}}{2}/\binom{n}{2}
}
{ \frac{1}{2}\left[\sum_j^r\binom{n_{.j}}{2}  + \sum_i^k\binom{n_{i.}}{2}	\right]
		-\sum_j^r \binom{n_{.j}}{2} \sum_i^k\binom{n_{i.}}{2}/\binom{n}{2}
} \;= ARI(U,V) \quad  \quad \qed\\
&\text{{\scriptsize $(*), (**)$ same as proof of identity 1.}  }
\end{align*}
On the other hand for the $NMI$, we have:
\begin{align*}
1 - \AG_{x\log x}^{|\cap|}(U,V)& =
\frac{	\sum\limits_{v\in V} \sum\limits_{u\in U} {n_{uv}}\log {n_{uv}}
- E(\sum\limits_{v\in V} \sum\limits_{u\in U} {n_{uv}}\log {n_{uv}} )
}
{ \frac{1}{2}\left[
			\sum\limits_{v\in V} {n_{.v}\log {n_{.v}}}
		  + \sum\limits_{u\in U} n_{u.}\log{n_{u.}}
		\right] - E(\sum\limits_{v\in V} \sum\limits_{u\in U}{n_{uv}} \log {n_{uv}})
}\\
where \; E(\sum\limits_{v\in V}& \sum\limits_{u\in U} {n_{uv}}\log {n_{uv}} ) = \sum_{u\in U}\sum_{v\in V}  \left(\frac{\eta_{.v}\eta_{u.}}{\sum_{u\in U}\sum_{v\in V} \eta_{uv} }\right) \log\left(\frac{\eta_{.v}\eta_{u.}}{\sum_{u\in U}\sum_{v\in V} \eta_{uv} }\right)
\end{align*}
\begin{align*}
\Rightarrow \;&1 - \AG_{x\log {x}}^{|\cap|}(U,V) \overset{*,**}{=}
\frac{	\sum_j^r \sum_i^k n_{ij}\log {n_{ij}}
		- \sum_i^k\sum_j^r \frac{n_{.j}n_{i.}}{n}\log {\frac{n_{.j}n_{i.}}{n}} 
}
{ \frac{1}{2}\left[\sum_j^r n_{.j}\log {n_{.j}}  + \sum_i^k n_{i.}\log {n_{i.}}\right]
		- \sum_i^k\sum_j^r \frac{n_{.j}n_{i.}}{n}\log {\frac{n_{.j}n_{i.}}{n}}}  
		\\
& = \frac{	n \sum_j^r \sum_i^k \frac{n_{ij}}{n}\log \frac{n_{ij}}{n} + n \log n
		- \sum_i^k\sum_j^r \frac{n_{.j}n_{i.}}{n} [ \log {\frac{n_{.j}}{n}}+\log {\frac{n_{i.}}{n}}+\log {n}] 
}
{ \frac{n}{2}\left[\sum_j^r \frac{n_{.j}}{n} \log {\frac{n_{.j}}{n}} + \sum_i^k \frac{n_{i.}}{n}\log \frac{n_{i.}}{n}  + 2\log n	\right]
		- \sum_i^k\sum_j^r \frac{n_{.j}n_{i.}}{n}[ \log {\frac{n_{.j}}{n}}+\log {\frac{n_{i.}}{n}}+\log {n}] }  
		\\
& = \frac{	-H(U,V) + \log n 
- \sum_i^k\frac{n_{i.}}{n}\sum_j^r \frac{n_{.j}}{n} \log {\frac{n_{.j}}{n}}+\sum_i^k\frac{n_{.j}}{n} -\sum_j^r \frac{n_{i.}}{n}\log {\frac{n_{i.}}{n}}-\sum_i^k\sum_j^r \frac{n_{.j}n_{i.}}{n^2}\log {n}
}{ \frac{1}{2}\left[-H(U) - H(V) \right]  + \log n	
		+ \sum_i^k\frac{n_{i.}}{n}H(V) +\sum_i^k\frac{n_{.j}}{n}H(U)-\log {n}  }\\
& = \frac{	- H(U,V) + H(V) +H(U)
}
{ -\frac{1}{2}\left[H(U) + H(V) \right] + H(V) +H(U) 	}
			= \frac{I(U,V)}{\frac{1}{2}\left[H(U) + H(V) \right]} = NMI_{sum}(U,V) \quad  \quad \qed\\
&\text{{\scriptsize $(*), (**)$ same as proof of identity 1.}  }\\
\end{align*}
\end{minipage}
\proofOf{\refthe{th:deltaeqRI}} 
\label{app:RIARI}
First we prove that in general cases we have:\vspace{-5pt}
\[\|UU^T - VV^T\|^2_F = \|U^TU\|^2_F + \|V^TV\|^2_F - 2\|U^TV\|^2_F \]
where $\|.\|^2_F$ is squared Frob norm. This holds since we have:
{\scriptsize
\begin{align*}\vspace{-5pt}
\|UU^T - VV^T\|^2_F &=  \sum_{ij}(UU^T - VV^T)_{ij}^2\\[-3pt]
& = \sum_{ij}(UU^T)_{ij}^2 + \sum_{ij}(VV^T)_{ij}^2 - 2\sum_{ij}(UU^T)_{ij}(VV^T)_{ij} \\[-3pt]
  &=  \|UU^T\|^2_F + \|VV^T\|^2_F - 2 |UU^T \circ VV^T| 
\end{align*}
}
Where the $\circ$ is element-wise matrix product, a.k.a. hadamard product, and $|.|$ is sum of all elements in the matrix\footnote{This equality is also useful in the implementation to improve the scalability.
}. The proof is complete with showing:
{\scriptsize
\begin{align*}  
|UU^T \circ VV^T| &= tr ((UU^T)^T VV^T)
=  tr (V^TUU^TV) =  tr ( (U^TV)^T U^TV) = ||U^TV||^2_F  
\\[-3pt]
||UU^T||^2_F &=tr((UU^T)^T UU^T) 
		=tr (U^TU U^TU) =tr ( (U^TU)^T U^TU) = ||U^TU||^2_F  
  \end{align*}
}
Now, we can prove \refthe{th:deltaeqRI} for the cases of disjoint hard clusters, using the notation, $n_{ij} = (U^TV)_{ij}$, we have  $\|U^TV\|^2_F = \sum_{ij} n^2_{ij}$ and:
\vspace{-2pt}{\scriptsize
\begin{align*} 
\|U^TU\|^2_F 
&=  \sum_{ij} <U_{.i},U_{.j}>^2 =  \sum_{ij} (\sum_k u_{ki}u_{kj})^2 
\overset{*}{=} \sum_{i} (\sum_k u^2_{ki})^2  \overset{**}{=} \sum_{i} (\sum_k u_{ki})^2  \overset{***}{=} \sum_{i} n_{i.}^2 \vspace{-15pt}
\end{align*}}
{
\scriptsize $(*)$ with assumption that clusters are disjoint, $ u_{ki}u_{kj}$ is only non-zero iff $i=j$ \\
\scriptsize $(**)$ with the assumption that memberships are hard, $u_{ki}$ is either $0$ or $1$, therefore $u_{ki}= u^2_{ki}$ \\
\scriptsize $(***)$ marginals of $N$ give cluster sizes in $U$ and $V$, i.e. 
$n_{i.} = \sum_{j} n_{ij} = \sum_{k} u_{ki}=|V_i| $ 
}
\hfill\\\hfill\\
Therefore for disjoint hard clusters we get:
\vspace{-5pt}
\begin{equation*}
\|UU^T - VV^T\|^2_F = \sum_{i} n_{i.}^2  +\sum_{j} n_{.j}^2  - 2 \sum_{ij} n^2_{ij}\vspace{-5pt}
\end{equation*}
The $RI$ normalization assumes that all pairs are in disagreement, i.e. $ NF_{RI} = |\mathbf{1}_{n\times n}| = n^2 $, as $max(max(UU^T), max(VV^T))=1$. The $ARI$ normalization compares $\Delta$ to the difference where the two random variable of $UU^T_{ij}$ and $VV^T_{ij}$ are independent, in which case we would have:
\begin{equation*}
 E( UU^T_{ij}VV^T_{ij}) = E((UU^T)_{ij}) E((VV^T)_{ij})\vspace{-5pt}
\end{equation*}
which is calculated by:\vspace{-3pt}
 {\small \begin{equation*} \frac{\sum_{ij} ((UU^T)_{ij} (VV^T)_{ij} )}{n^2} = \frac{\sum_{ij} (UU^T)_{ij}}{n^2} \frac{\sum_{ij} (VV^T)_{ij}}{n^2}\vspace{-3pt}
 \end{equation*}}
   Since 
 \(\Delta = ||UU^T - VV^T||^2_F = ||UU^T||^2_F + ||VV^T||^2_F - 2Sum (UU^T \circ VV^T )\), 
we have $ARI = 0$ or $\Delta/NF_{ARI} = 1$, i.e. agreement no better than chance, when this independence condition holds, i.e.:\vspace{-5pt}
 \begin{equation*} \Delta = NF_{ARI} \iff Sum (UU^T \circ VV^T ) = \frac{|UU^T||VV^T|}{n^2} \end{equation*}
\end{document}